\documentstyle[aaspp4,flushrt]{article}
\input psfig.sty

\newcommand\lsim{\mathrel{\rlap{\lower4pt\hbox{\hskip1pt$\sim$}}
        \raise1pt\hbox{$<$}}}
\newcommand\gsim{\mathrel{\rlap{\lower4pt\hbox{\hskip1pt$\sim$}}
        \raise1pt\hbox{$>$}}}

\def\simlt{\mathrel{\spose{\lower 3pt\hbox{$\mathchar''218$}}
     \raise 2.0pt\hbox{$\mathchar''13C$}}}
\def\simgt{\mathrel{\spose{\lower 3pt\hbox{$\mathchar''218$}}
     \raise 2.0pt\hbox{$\mathchar''13E$}}}

\def\thetaS{\mbox{\boldmath$\theta_S$}}
\def\thetaI{\mbox{\boldmath$\theta_I$}}
\def\thetaD{\mbox{\boldmath$\theta_D$}}
\def\angle0{\mbox{\boldmath$\theta_{\rm planet}$}}
\def\starangle{\mbox{\boldmath$\theta_{*}$}}
\def\thetaimp{\mbox{$\theta_{\rm tr. imp.}$}}
\def\gammaimp{\mbox{$\gamma_{\rm tr. imp.}$}}

\def\Deltabeta{\mbox{\boldmath$\Deltabeta$}}
\def\nablabf{\mbox{\boldmath$\nabla$}}
\def\nablaperp{\mbox{\boldmath$\nabla_\perp$}}
\def\Deltaell{\mbox{\boldmath$\Delta \ell$}}
\def\ellhat{\mbox{\boldmath$\hat \ell$}}

\begin{document}

\title{Atmospheric Lensing and Oblateness Effects \\During
an Extrasolar Planetary Transit}
\author{Lam Hui$^{1,2}$ and Sara Seager$^{2}$\\
\vspace{0.2cm}
$^{1}$ Department of Physics, Columbia University, 538 West 120th Street,
New York, NY 10027\\
$^{2}$ Institute for Advanced Study, School of Natural Sciences, Einstein
Drive,
Princeton, NJ 08540\\
{\tt lhui@astro.columbia.edu, seager@ias.edu}}

\begin{abstract}
Future high precision photometric measurements of transiting
extrasolar planets promise to tell us much about the characteristics
of these systems.  We examine how atmospheric lensing and (projected)
planet oblateness/ellipticity modify transit light curves.  The large
density gradients expected in planet atmospheres can offset the
unfavorably large observer-lens to source-lens distance ratio, and
allow the existence of caustics.  Under such conditions of strong
lensing, which we quantify with an analytic expression, starlight from
all points in the planet's shadow is refracted into view producing a
characteristic slowing down of the dimming at ingress (vice versa for
egress). A search over several parameters, such as the limb darkening
profile, the planet radius, the transit speed, and the transit geometry, 
cannot
produce a nonlensed transit light curve that can mimic a lensed light
curve. The fractional change in the diminution of starlight is
approximately the ratio of atmospheric scale height to planet radius,
expected to be $1 \%$ or less. The lensing signal varies strongly with
wavelength---caustics are hidden at wavebands where absorption and
scattering is
strong.  Planet oblateness induces an asymmetry to the transit light curve
about the point of minimum flux, which varies with the planet
orientation with respect to the direction of motion. The fractional
asymmetry is at the level of $0.5 \%$ for a projected oblateness of
$10 \%$, independent of whether or not lensing is important.  For
favorable ratios of planet radius to stellar radius (i.e. gas giant
planets), the above effects are potentially observable with future
space-based missions.  Such measurements could constrain the planet
shape, and its atmospheric scale height, density and refractive coefficient,
providing information on its rotation, temperature and composition.  For
HD~209458b, the only currently known transiting extrasolar 
planet, caustics are
absent because of the very small lens-source
separation (and a large scale height caused by a high temperature from the 
small separation). Its oblateness is also expected to be small because
of the tidal locking of its rotation to orbital motion.
Finally, we provide estimates of other variations to
transit light curves that could be of comparable importance ---
including rings, satellites, stellar oscillations, star spots, and
weather.
\end{abstract}

\section{Introduction}
\label{intro}

The exciting discovery of the transit of HD~209458 (Charbonneau et
al. 2000a, Henry et al. 2000, Jha et al. 2000) allowed the first direct
measurement of the physical properties of an extrasolar planet, including
its radius, and together with radial velocity measurements, its
absolute mass and average density.
Photometric followup of radial velocity planet candidates, and
future ground- and space-based surveys are expected to discover
hundreds more transiting planets.  Planned or proposed space-based
missions, in particular (MOST, MONS, COROT, the Eddington Telescope,
the Kepler Mission), are expected to be able to detect flux variations
at the $10^{-5}$ level. Recent Hubble Space Telescope observations of
the HD~209458b transit (Brown et al. 2001), with an accuracy of
$10^{-4}$, illustrate the capability of high precision space based
photometry. With the anticipated large number of transiting planets,
and the high accuracy with which they can be monitored, it is
important to explore small variations in the transit light curve that
might be detectable and allow us to deduce further properties of the
transiting planets.

In this paper we focus on how atmospheric refraction, which
we also refer to as atmospheric lensing or lensing for short,
modifies the transit light curve. We treat the general case
of an ellipsoidal planet, and the same calculation also
allows us to examine the effects of (projected) planet
oblateness/ellipticity,
regardless of whether or not lensing is
important. We examine under what conditions
atmospheric lensing and oblateness signatures might be significant,
detectable, and distinguishable from other effects.

We develop the lensing formalism in \S \ref{lensingmodel}, and
contrast atmospheric lensing with gravitational lensing. For readers
with a background in the latter, we point out several places where
intuition gained from gravitational lensing fails for atmospheric
lensing. In \S \ref{parameter} we give a complete list of all the
parameters employed in our model.  The condition under which atmospheric
lensing produces strong lensing (i.e. existence of caustics) is taken up in
\S
\ref{stronglensing}.  An analytic expression describing this condition
is given.
We emphasize the connections between
atmospheric lensing by extrasolar planets and atmospheric lensing by solar
system
bodies during their occultations of distant stars; the latter has long
been observed (see e.g. Elliot \& Olkin 1996; Hubbard 1997, and references
therein).
The difference is: in the extrasolar case, the lens-source distance is much
smaller than the lens-observer distance, and the source is extended
compared to the lens, while the opposite is true in the solar system case.
In \S \ref{explore}, we examine
how the lensing modifications to the light curve vary with the model
parameters. In \S \ref{signature}, we address the question: to what
extent can a lensed light curve be confused with a nonlensed light
curve with different parameters? Similarly, in \S \ref{ellipticity} we
isolate characteristic modifications by the planet's oblateness to the
transit light curve.  The issue of how absorption affects the color
dependence of lensing is taken up in \S \ref{absorption}, where we
also discuss how lensing impacts the planet transmission spectrum.
Finally, we conclude in \S \ref{discuss} with a summary and with a
list of other effects that might be of comparable magnitude
to the lensing and oblateness signatures.

An interesting paper by Hubbard et al. (2001) recently appeared
as our paper was nearing completion. They discussed some of the same
issues addressed here, but focused mainly on detailed predictions for
the case of HD~209458b. They also presented calculations of
atmospheric glow from Rayleigh scattering which we do not treat in
this paper, although we do discuss extinction by Rayleigh scattering.

\section{The Lensing Model}
\label{lensingmodel}

\subsection{Formalism}
\label{formalism}

Much of the formalism presented here
closely parallels that used for gravitational lensing, and the
similarities and differences are discussed in \S\ref{aVSg}.
Equivalent techniques have been applied to lensing of distant
stars by solar system bodies (Elliot \& Olkin 1996; see also
Draine 1998 for a treatment of gaseous lensing by large spherical clouds).
Ray optics
provides the following starting point for lensing:
\begin{equation}
\thetaS = \thetaI + \thetaD {D_{\rm LS} \over {D_{\rm OL} + D_{\rm LS}}},
\label{raytrace}
\end{equation}
where $\thetaD$ is the angle of deflection, $\thetaS$ and $\thetaI$
are the source and image positions, $D_{\rm LS}$ denotes the distance
between the lens and the source, and $D_{\rm OL}$ is the separation
between the lens and the observer.  The angle of deflection is
determined by spatial gradient of the refractive index $n$:
\begin{eqnarray}
\label{thetaD}
\thetaD = - \sum \nablabf n \times \Deltaell \times \ellhat,
\end{eqnarray}
where the sum is over individual segments of the ray $\Deltaell$ with
$\ellhat$ as the unit vector, and where $\nablabf$ is the spatial
gradient.

For atmospheric lensing,
$n = 1 + \alpha \rho$ where $\rho$ is the gas density and $\alpha$ is
a refractive coefficient that depends on both gas composition and
wavelength.
A common parametrization, known as Cauchy's formula, gives
$\alpha \rho = A_1 (1 + B_1/\lambda^2)$ where
$\lambda$ is the wavelength, and $A_1$, $B_1$ for common gases are given in
Table
\ref{alphatab} (Born \& Wolf 1999 p. 101).
For a ${\rm H_2 - He}$ mixture with $24\%$ He by mass,
$\alpha = 1.243 {\,\rm cm^3 g^{-1}}$ at $4400$~\AA, and
$\alpha = 1.214 {\, \rm cm^3 g^{-1}}$ at $6700$~\AA.

We are interested in cases where the deflection angle is small, and the lens
is thin (i.e. distances over which significant deflection occurs is small
compared to $D_{\rm LS}$ and $D_{\rm OL}$). Suppose the axis connecting the
observer
and (the center of) the lens points in the z direction. We have
\begin{equation}
\thetaD = \nablaperp \int_{-\infty}^\infty n dz = \nablaperp
\int_{-\infty}^\infty
 \alpha \rho dz,
\label{thetaD1}
\end{equation}
where $\nablaperp$ is the spatial gradient in the x and y directions.
We use ${\bf b} = (b_1 , b_2)$ to denote the impact vector---the
vector in x-y plane from the lens center to the point of impact.

The inverse of the magnification matrix is
\begin{eqnarray}
\label{Aij}
A^{-1}_{ij} & \equiv & {\partial \theta_S^i \over \partial \theta_I^j} \\
\nonumber
& = & \delta_{ij} + {D_{\rm OL} D_{\rm LS} \over
D_{\rm OL} + D_{\rm LS}} {\partial^2 \over \partial b_j \partial b_i}
\int_{-\infty}^\infty
 \alpha \rho dz,
\end{eqnarray}
where we have used ${\bf b} = D_{\rm OL} \thetaI$.

We are interested in a density profile $\rho$ which is ellipsoidal
in general, in
the sense that $\rho = \rho (r)$ where $r^2 = {x'}^2 / a_1^2 + {y'}^2 /
a_2^2 + {z'}^2 / a_3^2$.  \footnote{A spherical planet is simply a special
case within the class of models we study here.}
The axes denoted by $x'$, $y'$, and $z'$ are
generally not lined up with the $x$, $y$, and $z$ axes defined before.  We
show in Appendix A that
\begin{itemize}
\item such a profile is well-motivated;
\item in the simple case of an isothermal atmosphere in hydrostatic
equilibrium, the density profile is $\rho (r) = \rho_0 {\rm exp}
[-(r-R_0)/H]$ where $R_0$ is a reference radius whose choice will be
fixed below, $\rho_0$ is the density at $R_0$, and $H=(k_{b} T / g \mu
m_H)$ is the atmospheric scale height where $k_b$ is Boltzmann's
constant, $T$ is the temperature, $g$ the surface gravity, $\mu$ the
mean molecular weight, and $m_H$ the mass of the hydrogen atom;
\item The quantity $r$, which we loosely refer to as ``radius'', can
be written as $r = $ $\sqrt{(1-\epsilon) b_1^2 + (1+\epsilon) b_2^2 +
z^2}$ by a suitable rotation of axes, along with rescaling of the
density field---as long as the deviation from spherical symmetry is
small: $\epsilon \ll 1$, or $a_1$, $a_2$ and $a_3$ are not too
different from each other. The projected oblateness $\epsilon$
is
related to the actual oblateness of the planet, $\epsilon_A \equiv (a_1 -
a_3) / a_1$ (assuming it is axially symmetric with $a_1 = a_2$), by
$\epsilon = \epsilon_A (1 - {\rm cos}^2 \beta)$ where $\beta$ is an
Euler rotation angle.\footnote{In this paper, we use the terms oblateness
and ellipticity
interchangeably.}
$\beta$ is the angle between the axis of
rotation of the planet and the line of sight (see Appendix A for
details). Assuming $\beta$ is randomly distributed implies $\epsilon =
\epsilon_A/2$ on average.
\end{itemize}

We also work under the simplification that $\alpha$, the
density-independent refractive coefficient, is independent of position
on the planet. This is a simplification because the atmospheric
composition---hence the net value of $\alpha$---is expected to vary
with atmospheric depth.

Putting the above together, and using $\nablaperp \rho(r) = (1 \mp
\epsilon) ({\bf b}/r) {\partial\rho/\partial r}$ (upper/lower sign for
$b_1$/$b_2$) in equation~(\ref{thetaD}), the lensing equation
(equation~(\ref{raytrace})) can be written as
\begin{eqnarray}
\label{lensmaster}
&& \theta_S^1 = \theta_I^1 + (1-\epsilon)\theta_I^1 \psi(u), \quad
\theta_S^2 = \theta_I^2 + (1+\epsilon)\theta_I^2 \psi(u)\\ \nonumber
&& {\rm where} \quad \psi(u) \equiv \alpha
{D_{\rm LS} D_{\rm OL} \over D_{\rm OL} + D_{\rm LS}}
\int_{-\infty}^\infty {\partial \rho \over \partial r}  {dz \over r} \\
\nonumber
&& \quad \quad \quad \quad \rho = \rho_0 {\,\rm exp}[-(r-R_0)/H] \\
\nonumber
&& \quad \quad \quad \quad r^2 = D_{\rm OL}^2 u^2 + z^2, \, \, \, \, u^2
\equiv
(1 - \epsilon) {\theta_I^1}^2 + (1+\epsilon) {\theta_I^2}^2.
\end{eqnarray}
With the above form, the problem of predicting image position(s) given
a source position can be reduced to solving a simple single variable
equation---this and associated computational tricks are discussed in
Appendix B.

The magnification is given by (equation~(\ref{Aij})):
\begin{eqnarray}
\label{A}
&& A \equiv {\,\rm det} A_{ij} = [1 + 2 \psi + (1-\epsilon^2) \psi^2
+ (1-\epsilon^2) u^2 \psi \tilde\psi + (u^2 + \epsilon v) \tilde\psi]^{-1}\\
\nonumber
&& {\,\rm where} \quad v \equiv -(1-\epsilon) {\theta_I^1}^2 + (1+\epsilon)
{\theta_I^2}^2
\\ \nonumber
&& \quad \quad \quad \quad u^2 \equiv (1-\epsilon) {\theta_I^1}^2 +
(1+\epsilon) {\theta_I^2}^2
\\ \nonumber
&& \quad \quad \quad \quad \tilde\psi(u) \equiv {1\over u} {\partial \psi
\over \partial u}
= \alpha D_{\rm OL}^2 {D_{\rm OL} D_{\rm LS} \over
D_{\rm OL} + D_{\rm LS}} \int_{-\infty}^\infty \left[ {1\over r^2}
{\partial^2 \rho
\over \partial r^2} - {1\over r^3} {\partial \rho\over \partial r} \right]
dz,
\end{eqnarray}
and where $u$, $\psi$ and $\rho$ are
as described in equation~(\ref{lensmaster}).
The caustic is defined by source positions where $A$ diverges. The
critical curve is the image of the caustic. We discuss in Appendix B
how to find both, if they exist---a situation referred to as strong
lensing.

Finally, the observed flux from a star
during an extrasolar planet transit is given by:
\begin{eqnarray}
\label{F}
F(t) = \int d^2\theta_S \sum A I(\thetaS - \starangle(t)) W(\thetaI)
\end{eqnarray}
where $I(\thetaS -\starangle(t))$ is the surface brightness of the
star as a function of source position. The symbol $\starangle (t)$
denotes the position of the star's center as a function of time.  We
have chosen the origin of $\thetaS$ and $\thetaI$ to be centered at
the planet. The sum is over all images for a given source
position. The kernel $W(\thetaI)$ describes the occultation and
absorption; in this paper, we focus on a simple model in which
$W(\thetaI)$ is a step function,
\begin{eqnarray}
\label{W}
W(\thetaI) &=& 1 \quad \quad {\rm if} \quad \quad \sqrt{(1-\epsilon)
{\theta_I^1}^2 + (1+\epsilon) {\theta_I^2}^2}
> R_0/D_{\rm OL} \\ \nonumber
&=& 0 \quad \quad {\rm otherwise}.
\end{eqnarray}
This step function specifies $R_0$. It is the (elliptic) radius
below which the projected density exceeds some value such that the
atmosphere becomes completely opaque, or it is the radius at the rocky
surface of a planet.  A more realistic treatment of absorption will
have $W$ changing more gradually than this step function, and also
changing with wavelength; this will be discussed in \S
\ref{absorption}. As we will see, our step function model is actually
a good approximation to reality.
Note that both $I$ and $A$ above are functions of
wavelength.  Finally, for most purposes, we are only interested in the
normalized $F(t)$ i.e. $F(t)$ (equation(\ref{F})) divided by its
asymptotic value well away from the transit --- the stellar flux: $\int
d^2 \theta_S I(\thetaS)$. From now on we use $F(t)$ to refer
to the normalized value, and whenever we refer to ``flux'' in
this paper, we always mean the stellar flux normalized by its
pre- or post-transit value.

To summarize, equations (\ref{lensmaster}), (\ref{A}), and (\ref{F})
completely specify the problem of atmospheric lensing during a
planetary transit, for an isothermal atmosphere in hydrostatic equilibrium
with an ellipsoidal density profile.

\subsection{Atmospheric versus Gravitational Lensing}
\label{aVSg}

Gravitational lensing can be described by essentially the same equations
presented
above, except that the refractive index $n$ in equation~(\ref{thetaD1})
is equal to $1- 2 \phi$, where $\phi$ is the gravitational potential,
instead of $1 + \alpha \rho$. That $\alpha$ is
wavelength dependent implies atmospheric lensing is color dependent whereas
gravitational lensing is achromatic.

Similarly, gravitational lensing by an elliptic potential obeys
equations (\ref{lensmaster}), (\ref{A}), and (\ref{F}) with
$\alpha\rho$ replaced by $-2\phi$.
It is interesting
to note, however, that a well known theorem in gravitational lensing,
the magnification theorem, does not hold in atmospheric lensing.  The
magnification theorem states that for a given source position, the
magnification of all images must sum to at least unity. This can be
traced back to the fact that $\nabla^2 \phi$ is proportional to the
mass density which is positive definite.  For atmospheric density, the
relevant quantity, $\nabla^2 \rho$, is not guaranteed to be positive
definite. In other words, in atmospheric lensing the magnification
need not sum to unity, and thus a net suppression of flux can occur.

Another important difference between gravitational and atmospheric
lensing is that gravitational lensing is almost never significant when
the lens is very close to the source or the observer. This arises from
the fact that the combination of distances $D_{\rm LS} D_{\rm OL} /
(D_{\rm OL} + D_{\rm LS})$ becomes small (dominated by
the smaller of the two distances) if $D_{\rm LS} \ll
D_{\rm OL}$ or $D_{\rm OL} \ll D_{\rm LS}$. While the same factor
applies to both types of lensing, atmospheric lensing has the
advantage of having available an exponential density field (or nearly
so: see Appendix A). The analog for gravitational lensing, an
exponential potential, almost never occurs in nature; power-law fall
off is far more prevalent.  An exponential density profile allows
large gradients to offset an unfavorable combination of
distances. To be more precise, while a power-law profile would
result in factors of $D_{\rm LS} D_{\rm OL} /
[(D_{\rm OL} + D_{\rm LS}) R_0]$ in the relevant expressions (e.g. $\psi(u)$
in
equation (\ref{lensmaster})), an exponential profile gives
$D_{\rm LS} D_{\rm OL} /
[(D_{\rm OL} + D_{\rm LS}) H]$, which is considerably larger
(because atmospheric scale height $H$ $\ll$ planet radius $R_0$).
Nature offers a nice existence proof: atmospheric lensing
of distant stars by the solar system planets have been observed,
in spite of the fact that $D_{\rm OL} \ll D_{\rm LS}$ (see \S \ref{solar}).
The extrasolar
case we are interested in can be viewed as the symmetrical opposite
with $D_{\rm LS} \ll D_{\rm OL}$.

\section{Parameter Accounting}
\label{parameter}

Several parameters enter into the problem of planetary atmospheric
refraction during a transit.  However, most of them appear
in a few combinations. It is helpful to list them explicitly.

The quantities $\psi (u)$ and $u^2 \tilde\psi (u)$ in the lens mapping
equation (\ref{lensmaster}) can be well approximated by the following
expressions if $u D_{\rm OL}/H \gg 1$. This is true in realistic cases
because $u D_{\rm OL}$ is constrained to be larger than $R_0$ in order
for an image not to be obscured (occulted or absorbed) by the planet.
\begin{eqnarray}
\label{psirewrite}
&& \psi (u) = - B \sqrt{\pi/2} (u D_{\rm OL}/H)^{-1/2} {\,\rm exp}[-(u
D_{\rm OL} -R_0)/H]
\\ \nonumber
&& u^2 \tilde \psi(u) = B \sqrt{\pi/2} (u D_{\rm OL}/H)^{1/2}
[1 + (u D_{\rm OL} /H)^{-1}] {\,\rm exp}[-(u D_{\rm OL} -R_0)/H]
\\ \nonumber
&&
B \equiv 2 \alpha {\rho_0 \over H} {D_{\rm LS} D_{\rm OL} \over
{D_{\rm OL} + D_{\rm LS}}}.
\end{eqnarray}

Note that refraction is important (or the magnification $A$ is
significantly different from unity) only if $u D_{\rm OL}$ is close to
$R_0$.  Therefore, as far as the gross lensing behavior (or lack
thereof) is concerned, only three parameters are important, $B$ as
defined above, $R_0/H$, and $\epsilon$.

In addition to the three lensing parameters, we have three transit
parameters. They are the transit impact parameter (as opposed to the
{\it lensing} impact parameter) $\thetaimp$ (the distance of closest
approach between the center of the planet and the center of the star
in angular units),\footnote{The parameter $\thetaimp$ is related to
the planet orbital inclination $i$ by $\thetaimp = [D_{\rm LS}/D_{\rm
OL}] {\rm cos} \, i$.}  the transit-impact-angle $\gammaimp$ (the
angle between the major axis of the projected planetary ellipse and
the direction of transit motion), and the transit velocity $w$ in angular
units per unit time.
See Figure \ref{geometry}
for an illustration.
Since $w$ basically rescales the time axis in
our prediction of the transit light curve, we do not treat it as a
free parameter in our predictions.  We estimate it by using $w =
\sqrt{G M_\odot / D_{\rm LS}} /D_{\rm OL}$.

Finally, there are the parameters that describe the star: the stellar
radius $R_*$, and the surface brightness of the star, or
its limb darkening function, which is often parametrized as:
\begin{eqnarray}
\label{limpdark}
I(\thetaS-\starangle) &=& 1 - u_{*} (1 - s) - v_* (1 - s^2) \quad {\rm
if} \quad |\thetaS - \starangle| D_{\rm OL} \le R_* \\ \nonumber &=& 0
\quad {\rm otherwise}, \\ \nonumber &{\rm where}& \quad s \equiv
\sqrt{1 - (|\thetaS - \starangle| D_{\rm OL}/R_*)^2},
\end{eqnarray}
and where $\starangle$ is the angular position of the star's center
(recall that we have chosen the origin to be at the center of the lens,
the planet).  The limb darkening parameters $u_*$ and $v_*$ are
generally wavelength dependent.  Typical values are $u_* = 0.8$, $v_*
= -0.225$ in red bands, $u_* = 0.99$, $v_* = -0.17$ in blue bands, and
$u_* = 0.93$, $v_* = -0.23$ in intermediate yellow bands (Cox 2000).
Clearly, as far as the gross features of the transit are concerned, it
is the ratio $R_*/R_0$ that is important, not the absolute size of
$R_*$ (other than a rescaling in the overall duration of the transit).

In summary, we have: three lensing parameters, $\epsilon$, $B$, and
$R_0/H$; two transit parameters, $\thetaimp$, and $\gammaimp$; and three
stellar parameters, ${R_*/R_0}$, $u_*$, and $v_*$. In the case of a
spherical planet the total number of parameters is reduced by two
because the lensing paramter $\epsilon$ and the transit parameter
$\gammaimp$ are not needed.  We will not perform an exhaustive study
of the full parameter space in this paper, but will be content with a
mostly qualitative (with a few exceptions) description of the
dependence of the transit light curve on these parameters.  Some of
these parameters are likely degenerate. For instance, changing $\thetaimp$
means the transit is sampling a different part of the stellar profile,
which might be mimicked by a different limb darkening law (e.g. a
different $u_*$ and $v_*$). Finally, note that for
extrasolar systems, $D_{\rm OL} \gg D_{\rm LS}$ and so
$D_{\rm OL}$ drops out of the lensing equation (e.g. equation (\ref{lensmaster})),
and appears only as an unobservable overall scaling of angular separations.

\section{Strong Lensing and Caustic Structure}
\label{stronglensing}

\subsection{Condition for Strong Lensing}
\label{strong}

The first question we would like to address is when strong
lensing---i.e. the existence of caustics or multiple images (which may
be unresolved)---occurs.  The existence of caustics could lead to
significant modifications in transit light curves.

Caustics in the source plane, or critical curves in the image
plane, can be obtained by solving for divergent magnification,
$A^{-1} = 0$ (equation~(\ref{A})). In the spherical case, with
$\epsilon = 0$, this is straightforward. Recall that images must have $u
> R_0/D_{\rm OL}$ to be visible, otherwise they are blocked (occulted
or absorbed) by the planet (equation~(\ref{W})). Therefore, the condition
for strong atmospheric lensing by a spherical planet is
\begin{equation}
1 + 2 \psi(u) + \psi(u)^2 + u^2 \psi(u) \tilde\psi(u) + u^2 \tilde
\psi(u) = [1 + \psi(u)] [1+\psi(u)+u^2 \tilde \psi(u)]< 0, \quad u =
R_0/D_{\rm OL}.
\end{equation}
This guarantees that some images with $u > R_0/D_{\rm OL}$ will have
$A^{-1} = 0$.  Using the results in \S \ref{parameter}, the above
condition imposes a relation between only two parameters: $B$ and
$R_0/H$.

Since $\epsilon$ is small, the condition for strong lensing by an
elliptic atmosphere will not be too different from the spherical case
described above.  We estimate it using results proved in Appendix B
(equation~(\ref{causticcond}) and the following paragraph), replacing
the above condition with the following:
\begin{equation}
1 + 2 \psi(u) + (1-\epsilon^2) \psi(u)^2
+(1-\epsilon^2) u^2 \psi(u) \tilde\psi(u)
+ (1-\epsilon) u^2 \tilde\psi(u) < 0, \quad u = R_0/D_{\rm OL}.
\end{equation}
Using equation~(\ref{psirewrite}),
this imposes a condition on three parameters, $\epsilon$, $B$ and $R_0/H$,
\begin{eqnarray}
\label{strongcondition}
1 - \sqrt{\pi \over 2} \sqrt{H \over R_0} B - \epsilon < 0,
\end{eqnarray}
where we have used $H/R_0 \ll 1$ and $\epsilon \ll 1$.
This is a main result of our paper.

The above condition for strong lensing is depicted in Figure
\ref{paramLENS}.  We show in the same figure the relevant parameters
for several known planets. We assume $\alpha = 1.2 {\,\rm cm^3/g}$ for
all of them.  Mars (M) has $R_0 \sim 3400$ km, $D_{\rm OL} D_{\rm LS}
/ (D_{\rm OL} + D_{\rm LS}) \sim 1.5$ AU, $\rho_0 \sim 1.5 \times
10^{-6} {\,\rm g/cm^3}$ and $H \sim 8.7$ km (Jones 1999 p. 314).
Jupiter (J) has $R_0 \sim 71000$ km, $D_{\rm OL} D_{\rm LS} / (D_{\rm
OL} + D_{\rm LS}) \sim 5.2$ AU, $\rho_0 \sim 3.5 \times 10^{-5} {\,\rm
g/cm^3}$ and $H \sim 21.7$ km (Jones 1999 p. 341).  The extrasolar
planet HD 209458b (HD) has
$R_0 \sim 10^5$ km, $D_{\rm OL} D_{\rm LS} / (D_{\rm OL} + D_{\rm LS})
\sim 0.05$ AU, $\rho_0 \sim 1 - 6 \times 10^{-6} {\,\rm g/cm^3}$ and
$H \sim 500 - 700$ km (Charbonneau et al. 2000, Henry et al. 2000,
Mazeh et al. 2000, Jha et al. 2000, Burrows et al. 2000).  The density
$\rho_0$ is obtained by computing the density at which the optical
depth for Rayleigh scattering is unity (at $\lambda = 6500$~\AA; see \S
\ref{absorption}),
or in the case of Mars, $\rho_0$ is the density at the base of its
atmosphere, just above its solid surface.
For a gaseous planet or a
rocky planet with a thick atmosphere, $\rho_0$ (and $R_0$) is
wavelength-dependent. For example, our own solar system planets have
much stronger molecular absorption in the UV and IR compared to the
optical. Strong absorption bands make $\rho_0$ smaller (where, again,
$\rho_0$ is the density and $R_0$ the radius at which the optical depth
$=1$), in some
cases to the point where atmospheric lensing is no longer strong (recall
that
$B \propto \rho_0$). This is illustrated by the point $J'$ where
we show Jupiter observed at a waveband where the absorption cross-section is
600 times larger e.g. Rayleigh scattering observed at
$\lambda = 1300$~\AA~instead of $6500$~\AA. The main effect is
a dramatic decrease in $B$. The increase in $R_0/H$ is very small,
because the radius varies only logarithmically with the
density. We will discuss the color dependence
of lensing in more detail in \S \ref{absorption}.

There is clearly a wide range of possibilities in the two-dimensional
parameter space of $B$ and $R_0/H$. J and HD, which have a similar
mass, occupy very different parts of the diagram primarily because
they are situated at very different distances from their parent
stars. Higher temperature in the latter leads to a much larger
atmospheric scale height. To investigate how variables such as stellar
luminosity, albedo, planet mass and composition control the
lensing behavior is outside the scope of this paper. In fact, the
radial velocity detections of extrasolar planets have taught us that
planet orbital characteristics can be quite different from
expectations based on the solar system. We therefore adopt the
philosophy that the parameters spanned by the above three examples are
all possible and interesting, and our goal is to understand the
importance of atmospheric lensing under these conditions. We will
briefly discuss the physical motivations for the choice of some
parameters in \S \ref{discuss}.

\subsection{Caustic Structure and Magnification}
\label{caustic}

The lower panel of Figure \ref{caustic2} shows the caustic (solid
line) and the critical curve (dotted line) for a fiducial model with
projected oblateness $\epsilon = 0.05$ denoted by the solid square in
Figure \ref{paramLENS} (see \S \ref{lightcurve} for details). A point
source situated within the caustic produces 4 images.\footnote{It is
interesting to note that lensing by an elliptic potential or density
profile generally produces two sets of caustic curves (e.g. Schneider,
Ehlers \& Falco 1992), whereas we see only one here. The origin can be
traced to the fact that $\psi (u)$ in equation (\ref{lensmaster}), which is
proportional to the deflection angle, is monotonic. We do not expect
small local deviations from the exponential density profile to change
this conclusion, because $\psi (u)$ involves an integral over many
layers of the atmosphere.}

The upper panel of Figure \ref{caustic2} shows the magnification as a
function of source position for a point source situated on the x-axis.
This figure shows that lensing has two effects. First, it
suppresses the flux originating from source positions just outside
$R_0$ ($u D_{\rm OL} > R_0$). Second, it brings in additional photons
from source positions behind $R_0$ ($u D_{\rm OL} < R_0$)---photons
which would otherwise be blocked
(occulted or absorbed) by the planet.  Which effect dominates can be
calculated by integrating over the star (regarded here as a collection
of point sources) and depends on details of the magnification profile
and the limb-darkening profile (i.e. the lensing parameters, transit
parameters, and stellar parameters described in \S\ref{parameter}). We
can derive a simple result from equation~(\ref{F}) for the limiting
case of a completely flat and constant $I$. Taking $I$ out of the
integral in equation~(\ref{F}), $F$ can be rewritten as $I \int
d^2 \theta_I W(\thetaI)$, which is identical to the result if no lensing
takes place.  It is not hard to see that the same conclusion follows
if $I$ is constant within some region, and if the planet is well
within this region.  In other words, if there is strictly no limb darkening,
there will be no net gain or loss of stellar flux from atmospheric lensing,
{\it provided} the projected planet is well within the stellar disk.

Realistic stellar profiles are never exactly constant. Refraction
therefore generally modifies the dimming or deficit of the stellar
flux during a transit, especially during ingress and egress. The sign
of the modification depends on the exact transit and lensing
parameters. The size of the modification can be crudely estimated
from the ratio of the projected area of the atmosphere to that of the
planet: $\sim$$H/R_0$, if strong
lensing occurs.

\subsection{Application to Solar System Occultations}
\label{solar}

Figure \ref{caustic2} also represents an occultation of a point-source
background star by a solar system planet. The dotted line shows what
one would expect for an atmosphereless planet (ignoring diffraction):
the stellar intensity is constant and vanishes instantaneously when
the star passes behind the planet as viewed by the observer.
The
dashed line shows the magnification from equation~(\ref A)
when refraction by the atmosphere is significant.
The magnification curve can be understood qualitatively
as follows. (1) The diminution of starlight during
ingress and egress is due to atmospheric differential refraction (not
by absorption which occurs at much lower depths in this model), which
causes stellar light rays to diverge away from the planet-observer
line-of-sight. This is a consequence of the radial density gradient in
the atmosphere.  (2) When the star is behind the opaque part of the
planet, the occultation light curve is not zero, because some of the
stellar light rays are refracted into the observer's line-of-sight.
(3) When the point source is near the geometric center of the planet,
the stellar flux is symmetrically refractively focused causing an
increase in brightness.  The solid line in Figure~\ref{caustic2} shows
the occultation light curve that includes the flux from the planet
from reflected sunlight if the occulted star has an intensity 20\% of
the planet. The solid curve shows that the planet's reflected light
dominates the light curve during the stellar occultation. Hence the
minimum observed brightness corresponds to the planet alone and the
full extent of refractive focusing and defocusing is not observed for
bright solar system planet occultations of background stars.  The
magnification peaks have been observed for most solar system planets
(and with the multiple images even resolved in the case of Saturn;
see Nicholson, McGhee, \& French, 1995).  Note that the magnification
peaks are generally reduced due to atmospheric absorption or scattering
and to the
finite size of the star.  Note also that if the star does not move
along the axes of symmetry of the solar system planet, the two
magnification peaks will generally be of different heights and widths.
Also, the two peaks will merge into one if the planet's oblateness is
sufficiently
small.
Refractive occultations of all of the solar system planets (except
Mercury) have been observed and in some cases well studied (see Elliot
\& Olkin 1996 and references therein). Models of them can provide
temperature ($T$), pressure ($P$), and density ($\rho$) as a function
of atmospheric depth.

\section{The Transit Light Curve}
\label{lightcurve}

\subsection{Exploring the $R_0/H - B$ Plane}
\label{explore}

Figure \ref{transit22} shows the transit light curve with or without lensing
for a
fiducial model denoted by the solid square in Figure \ref{paramLENS}: $B
= 40.3$, $R_0/H = 117.3$, $\epsilon = 0.05$, together with $R_0/R_* =
0.084$, $R_* / \sqrt{G M_\odot/D_{\rm LS}} = 233.66$ minutes, $u_* =
0.8$ and $v_* = -0.225$, $\thetaimp = 2 R_0 / D_{\rm OL}$ and
$\gammaimp = 45^0$. The last six parameters define the overall
light deficit and rough duration of transit, the limb darkening
profile, and the geometry of the transit. We emphasize that a
different value for $R_* / \sqrt{G M_\odot/D_{\rm LS}}$ (\S
\ref{parameter}) can be easily accommodated by rescaling the time axis
of all light curves shown below.

To better isolate the effect of lensing, we define and show in
Figure \ref{dtransit22} the following quantity (solid line):
\begin{equation}
f(t) = [\Delta F_{\rm lens} (t)- \Delta F_{\rm no \, lens} (t)]/
\Delta {F_{\rm lens}^{\rm max.}},
\label{ft}
\end{equation}
where $\Delta F(t) \equiv 1 - F(t)$ represents the light deficit, and
the subscripts denote whether or not lensing is taken into account
(with all other parameters fixed). We choose to normalize by the
maximum deficit, which can be estimated roughly by
$(R_0/R_*)^2$.\footnote{We could normalize by the maximum deficit in
the nonlensed case, but it would only modify $f(t)$ to higher order
($O(f(t)^2)$).}  The quantity $f$ is therefore a {\it fractional}
difference.  This gives us a nice way to estimate the {\it absolute}
size of the deficit for a transit with any given $R_0/R_*$: it is
roughly $(R_0/R_*)^2 (1 + f)$.  As we have discussed in \S
\ref{stronglensing}, $f$ can be estimated by $\sim H/R_0$, which is
around $1 \%$ for our fiducial model.  This is in rough
agreement with the results of our numerical integration.
The {\it absolute} change in flux
due to lensing is therefore $1 \% \times (R_0/R_*)^2$, which for the
example shown with $R_0/R_* \sim 0.1$, is $10^{-4}$---a potentially
significant effect.

The solid line in Figure \ref{dtransit22.5} shows the same quantity
$f$ for a transiting planet with the same $H/R_0$, but a
smaller $B$ ($= 14.9$)---denoted by an open square in Figure
\ref{paramLENS}. The overall magnitude of $f$ is a little smaller but not
too
dissimilar from the fiducial model above.
The sign of the lensing effect as the transit
progresses, however, is quite different.
As we have previously
discussed, two opposing lensing effects are competing.  One is the
suppression of flux from source positions just above $R_0$. The other
is the addition of photons from source positions in the planet's
shadow. For the model with a larger $B$, the first dominates over the
second effect at all times. For the model here with a smaller $B$, the
second effect dominates, at least temporarily towards the end of
ingress or beginning of egress.
At mid-transit, however, the first effect still dominates.
Interestingly, a search through a whole range of limb-darkening
parameters reveals that lensing almost always causes a net suppression
of flux at mid-transit, except when the stellar profile is very spikey
at the center.

When $B$ is decreased further so that caustics no longer exist, the
first effect, suppression of flux, will dominate.  This is illustrated
by the dotted line in Figure \ref{dtransit22.5}, which is the model
denoted by a cross in Figure \ref{paramLENS} and is chosen to resemble
HD~209458b (see \S \ref{strong}).\footnote{The projected
ellipticity is chosen to be $10^{-3}$,
although its precise magnitude does not affect our conclusions here.
The small oblateness is expected due to the tidal locking of
rotational and orbital motions, given the proximity of HD 209458b to
its companion star (see Hui \& Seager, in preparation).}
Clearly,
atmospheric lensing is weak for the close-in extrasolar giant planets
like HD 209458b. This is no surprise since, as we have pointed out in
\S \ref{strong} and Figure \ref{paramLENS}, there are no caustics in
the case of HD 209458b.  This is due to the low $B$ which is
mainly caused by a small lens-source distance
because $B$ is proportional
to $D_{\rm LS}$. A second, less direct effect of the small
$D_{\rm LS}$ is that the planet's temperature is hotter and
thus $H$ is higher, further decreasing $B$.
However, it is interesting that even though the
combination of parameters does not allow the existence of caustics in
the case of HD 209458b, refraction nonetheless could modify the light
curve at a (fractional) level of $\sim 5 \times 10^{-4}$ (depending on
the actual atmospheric structure; see \S\ref{discuss} for
details). Note also that because $D_{\rm LS} = 0.05$ AU for this
close-in gas giant (compared to $D_{\rm LS} = 1$ AU we adopt in
previous cases); the shorter orbital radius leads to a higher velocity
(the parameter $w$; see \S \ref{parameter}), hence a shorter transit.
Finally, the dashed line in Figure \ref{dtransit22.5} $f(t)$ shows the
model shown by a triangle in Figure \ref{paramLENS}.  The behavior is
similar to the solid line of Figure \ref{dtransit22}, except that the
overall effect is weaker. This is due to a larger $R_0/H$, $\sim 700$,
from which one expects $f$ of the order of $\sim 1.5 \times 10^{-3}$,
in agreement with what we find.

\subsection{The Lensing Signal}
\label{signature}

The previous section shows that lensing can modify the transit light
curve to an extent that is potentially detectable in some cases. If so, high
precision photometric observations of extrasolar planet transits can
provide planet atmosphere parameters such as scale height and density,
both of which affect the lensing behavior.  What is not clear,
however, is whether a transit light curve from a planet with lensing
can be distinguished from a transit light curve from a planet with
slightly different parameters, but with no lensing. This is what we
investigate here.

As shown in the last section, the modification introduced by
atmospheric lensing is qualitatively quite different for models with
or without caustics. For models with caustics, such as those shown in
Figure \ref{dtransit22}, $f(t)$ experiences a significant dip in the
light deficit around ingress and egress (the same holds for solid and
dashed lines in Figure \ref{dtransit22.5}).  This dip is caused by the
additional photons brought in from the planet's shadow due to lensing
which causes the light deficit to drop temporarily.  In other words,
the inexorable dimming of star light at ingress is temporarily slowed
down by the additional photons that are refracted into view.  This
effect is much weakened if caustics do not exist, as in the case of
the dotted line in Figure \ref{dtransit22.5}, which is our model for
HD 209458b.  This dip, or slight reduction in the rate of dimming,
introduces a distinct shape to the transit light curve during ingress
and egress, which one might hope to observe.

The dotted line in Figure \ref{dtransit22} shows the following
fractional difference:
\begin{equation}
\tilde f (t) = [\Delta F_{\rm lens} (t, P) - \Delta F_{\rm no lens} (t, P')]
/\Delta F_{\rm lens}^{\rm max} (P),
\label{ftilde}
\end{equation}
where the extra argument $P$ denotes the whole set of parameters that
determine the transit characteristics. We use $P'$ for the second term
(which does not take into account lensing)
to emphasize that the second term has a different set of parameters from the
first term. The question is how small one can make $\tilde f$ by
choosing $P'$ appropriately. If $\tilde f(t)$ is too small to be detectable,
then
for practical purposes, one cannot tell from the observed
lightcurve whether or not refraction plays a role, and therefore
cannot determine useful parameters from refraction effects.

The dotted line in Figure \ref{dtransit22} shows the result of our
effort of searching for the appropriate $\Delta F_{\rm no lens} (t,
P')$ that would minimize $\tilde f(t)$ for our fiducial model (solid
square in Figure \ref{paramLENS}).  We systematically vary a host of
parameters, including $R_0$, $u_*$, $v_*$, $w$, $\thetaimp$ and
$\gammaimp$.  The minimum fractional difference we can come up with is
a model that brings $\tilde f(t)$ down to about $10^{-3}$ during
mid-transit, but still almost $5 \times 10^{-3}$ at ingress and
egress.  It should be emphasized, however, that it is conceivable a
different limb darkening law from the one we have adopted
(equation~(\ref{limpdark})) could be chosen to further decrease
$\tilde f$.  Nonetheless, within the range of simple models we
consider, it appears that the lensing signal (i.e. the caustic signal)
cannot be completely masked by a clever choice of parameters.  A
discrepancy in $\tilde f$ of $5 \times 10^{-3}$ is potentially
observable; this {\it fractional} difference of $5 \times 10^{-3}$ means,
for a transit where $R_0^2 / R_*^2 \sim 10^{-2}$ say, the {\it
absolute} difference in the light deficit would be $\sim 0.5 \times
10^{-4}$.  We will discuss other nonlensing effects that might be of
similar magnitude in \S \ref{discuss}.

We find that the same conclusion holds for
other cases with caustics, such as those shown as
solid and dashed lines in Figure \ref{dtransit22.5}.
For cases without caustics, on the other hand, such as
that shown with a dotted line in Figure \ref{dtransit22.5},
we find that a suitable choice of $R_0$ for the second term
in equation~(\ref{ftilde}) is generally sufficient to
make $\tilde f$ quite small and undetectable.

The time derivative $dF/dt$ provides another way to illustrate the
effect of refraction.  Figure \ref{transitDnew22} shows the time
derivative $dF/dt$ for our fiducial model, with (solid line) or
without (dotted line) lensing. The bottom panel focuses on the part of
ingress where $dF/dt$ reaches its minimum value before increasing.
This corresponds roughly to when half of the planet has crossed the
stellar limb. What is interesting is that the presence of caustics
introduces a distinct enhancement in $dF/dt$ (solid over dotted curve)
around $t = -230$ min.  This is due to the additional photons brought
about by the caustic. From Figure \ref{caustic2}, we can see
that the caustic has a size of about $0.2 R_0 / D_{\rm OL}$.  Dividing
it by $w$, the angular velocity, we obtain $\sim 5 $ minutes, in rough
agreement with the duration of the enhancement seen in Figure
\ref{transitDnew22}.  Since the size of the caustic scales with
$\epsilon$, this offers a direct way of measuring the
projected oblateness of the planet. However, it appears that one needs to be
able to detect differences in $dF/dt$ at the level of at least one
part per million per minute, which is likely difficult to achieve in
the near future. Therefore, while the derivative $dF/dt$ offers an
interesting way to look at the effect of lensing, the integral
fractional differences $\tilde f$ or $f$ might be a more practical
probe.


\subsection{The Oblateness Signal}
\label{ellipticity}

The projected oblateness or ellipticity of a planet can affect the transit
light curve.  Here we explore the fractional difference
\begin{equation}
f_e (t) = [\Delta F(t,\epsilon =0.1) - \Delta F(t,\epsilon =0)]
/\Delta F^{\rm max} (\epsilon =0.1)
\label{fe}
\end{equation}
to determine the magnitude of the effect.
Here $\Delta F$ can either have lensing
taken into account or not because we would like to investigate how
oblateness affects the light curve irrespective of whether or not refraction
is present.
Figure \ref{diff33} shows $f_e$ for the fiducial model (solid square
in Figure \ref{paramLENS}). The solid line denotes a case
with lensing, and the dotted line without.
It is interesting that oblateness
introduces a modification to the light curve that is quite
similar in the two cases.

Next we pursue the same exercise as before and ask if parameters $P'$
can be found such that the following can be minimized:
\begin{equation}
\tilde f_e (t) = [\Delta F(t, \epsilon =0.1, P) - \Delta F(t, \epsilon =0,
P')]
/\Delta F^{\rm max} (\epsilon =0.1, P),
\label{fetilde}
\end{equation}
where $P$ denotes the parameters for our fiducial model. Figure
\ref{diff34.Ch} shows the results. We cannot find combinations of
parameters $P'$ to reduce $\tilde f_e$ below $10^{-3}$ at ingress and
egress, suggesting that we might be able to distinguish light curves
caused by ellipsoidal versus spherical planets.
The $P'$ that minimizes $\tilde f_e$, as shown in Figure \ref{diff34.Ch},
turns out to correspond to a spherical planet with slightly larger $R_0$,
and essentially the same area as the ellipse in the fiducial model.
In other words, the curves in Figure~\ref{diff34.Ch} basically show the
fractional
difference between the light deficit of a spherical planet and an
oblate planet ($\epsilon = 0.1$) with the same area (with or without
lensing).
This is a detectable signature.

How large is the lensing signature
compared to the oblateness signature for the same stellar parameters?
For our fiducial planet model, a comparison between
Fig. \ref{dtransit22} and (dotted line of) Fig. \ref{diff34.Ch} shows that
the two effects are of a comparable magnitude.
Note, however, the oblateness signature can persist
even if the parameters are not right for strong lensing.

The reader might have noticed that some of the previous light curves
are not symmetric around $t = 0$ (which is chosen to be
the point of mid-transit or minimum flux). This is due to the fact
that the major/minor axis of the planet is misaligned with the direction of
motion (see Figure \ref{geometry}) i.e. only $\gammaimp = 0$ or $\gammaimp =
90^0$ would
produce a light curve that is symmetric around $t = 0$.
In the presence of a general misalignment, the asymmetry
provides another useful diagnostic of oblateness
because a spherical planet can only produce symmetric lightcurves.
We show in Figure \ref{asymm} the following fractional difference,
\begin{equation}
f_{\rm asym} (t) = [\Delta F (t) - \Delta F(-t)] /\Delta F^{\rm max}.
\label{fasym}
\end{equation}
The solid line denotes the case for $\epsilon = 0.1$, and the dashed
line for $\epsilon = 0.05$, both for $\gammaimp = 45^0$.  The dotted
line shows the same for $\gammaimp = 0$, verifying that the asymmetry
vanishes for exact alignment of major (or minor) axis with the
direction of motion. The lower panel shows $f_{\rm asym} (t)$ where
all quantities are evaluated with lensing, while the upper panel shows
the same without lensing. The sets of curves are almost identical.
This means the degree of asymmetry in the light curve is determined
by the size of the oblateness and angle $\gammaimp$ alone, and is
insensitive to lensing.

\section{Absorption and Color Dependence}
\label{absorption}

\subsection{More Accurate Modeling of Absorption and its Effects on Lensing}

So far we have been using a crude step function model of absorption
that turns abruptly on and off at $u = R_0/D_{\rm OL}$.  Here we use
absorption to refer to extinction by either absorption or scattering.
Taking into account absorption in a more realistic way is in principle
straightforward. For each image position, there is an associated
optical depth:
\begin{eqnarray}
\label{tau}
\tau(\thetaI) &=& \int_{-\infty}^{\infty} \sigma {\rho \over \mu m_H} dz
\\ \nonumber
&=& {\rho_0 \over \mu m_H} \sigma \sqrt{2 \pi u D_{\rm OL} H} {\, \rm exp}
[-(u D_{\rm OL}
- R_0)/H],
\end{eqnarray}
where $u$ is defined in terms of $\thetaI$ as in equation~(\ref{A}), and we
have used the condition that $u D_{\rm OL}/H \gg 1$. The symbol $\mu
m_H$ denotes the mean molecular weight.  To compute the
light curve with absorption, $W(\thetaI)$ should be replaced by
$e^{-\tau(\thetaI)}$ in equation (\ref{F}).

The absorption cross-section $\sigma$ can vary significantly depending
on the composition of the planet atmosphere and the wavelength of
interest, and is essentially infinite at the surface of a rocky
planet.  To demonstrate concretely the effect of more accurate,
gradual absorption, we adopt the cross-section appropriate for
Rayleigh scattering (Jackson 1999):
\begin{equation}
\sigma = 10^{-27} {\,\rm cm^{2}} (5000 \AA / \lambda)^4,
\label{sigmaRay}
\end{equation}
where $\lambda$ is the wavelength of interest.

We recompute predictions for the fiducial model (solid square
in Figure \ref{paramLENS}) using the model outlined above.
In Figure \ref{diffabsorb}, we show the following fractional difference,
\begin{equation}
f_a (t) = [\Delta F_{\rm ab} (t) - \Delta F (t)]/\Delta F_{\rm ab}^{\rm max}
\label{fat}
\end{equation}
where $\Delta F_{\rm ab} (t)$ represents the deficit computed when
a gradual change in absorption is used, while $\Delta F (t)$ denotes
the same computed with a step function approximation for $W$
(equation~(\ref{W})).
The upper solid and dotted lines in Figure \ref{diffabsorb} give the
above quantity with or without lensing.
Gradual absorption produces light curves that are different from
a step function absorption/occultation at about the $1 \%$ level.
The above uses the same planet radius $R_0$ for
$\Delta F_{\rm ab}$ and $\Delta F$.
This difference can be much diminished, however, if one chooses an
appropriate $R_0$ for the step function model. The lower solid and
dotted lines in Figure \ref{diffabsorb} are the minimized fractional
difference:
\begin{equation}
\tilde f_a (t) = [\Delta F_{\rm ab} (t, R_0) - \Delta F (t, R_0')]/
\Delta F_{\rm ab}^{\rm max} (R_0),
\label{fatilde}
\end{equation}
for cases with and without lensing respectively, where we tune
$R_0'$ to make $\tilde f_a$ small. The small value of $\tilde f_a (t)$
in Figure \ref{diffabsorb}
shows that our step function
absorption model is actually a good approximation to reality---a
suitable $R_0$ can always be found such that it approximates a gradual
absorption model to high accuracy. This works primarily
because $e^{-\tau}$ does behave almost like a step function,
due to the fact that $\rho$ and therefore $\tau$ varies exponentially
with radius.

With the above results we can proceed to discuss the color dependence
of a transit light curve.  There are three main factors.  First, the
stellar profile changes with wavelength, generally flatter as one
considers redder wavebands (see \S \ref{parameter}).  Second, the
refractive coefficient $\alpha$ varies with wavelength.  The variation
depends on atmosphere composition, and some examples are given in
Table \ref{alphatab}.  Third, the absorption cross-section $\sigma$
also varies with wavelength.  In fact, for Rayleigh scattering for
instance (equation~(\ref{sigmaRay})), absorption varies much more
strongly with wavelength compared to both the index of refraction and
limb darkening.  This can have a dramatic consequence for the
existence of caustics. Consider a model like the open square in the $B
- R_0/H$ plane as shown in Figure \ref{paramLENS}. Recall that $B =
[2\alpha \rho_0 /H] [D_{\rm OL} D_{\rm LS} / (D_{\rm OL} + D_{\rm
LS})]$ (equation~(\ref{psirewrite})). As one considers shorter
wavelengths or bluer colors, $\sigma$ becomes larger, increasing the
optical depth (equation~(\ref{tau})).  In our step function model,
this is equivalent to lowering $\rho_0$ (defined to be the density at
which $\tau = 1$), in other words lowering $B$.\footnote{Changing
$\sigma$ also changes our definition of $R_0$ which is where $\tau
\sim 1$. While this does affect the transit light curve somewhat, its
effect on lensing is smaller compared with that due to $\rho_0$. This
is because varying $\sigma$ typically changes $R_0$ by a few scale
heights $H$, and that represents a small change to $R_0/H$ (since $R_0
\gg H$), which is the other parameter that controls the lensing
behavior.}  For sufficiently blue colors, the model would shift down
from the open square in Figure \ref{paramLENS} and cross the strong
lensing threshold (the solid line), erasing the caustics, and making
the lensing signal much weaker (as discussed in \S
\ref{signature}).
The same also holds true at wavebands where other kinds of absorption
are important, such as molecular electronic absorption bands in the UV,
and rotational-vibrational bands in the IR of species such as
H$_2$O, CO$_2$, and CH$_4$.
This is a well known result for solar system planet
occultations of distant stars where lensing effects are strong in the
optical but nonexistent in the UV and within strong absorption bands
in the IR.

In summary, for extrasolar planet transits where caustics exist, we
expect the lensing signal to disappear, or at least
weaken, at wavebands with high extinction.

\subsection{Planetary Atmosphere Transmission Spectrum and Stellar Spectrum}

Extrasolar planet transit transmission spectra have been described in
several papers
for both the close-in extrasolar giant planets (Seager \& Sasselov 2000;
Brown,
2001; Hubbard et al. 2001) and Earth-like planets (Schneider 1994;
Webb \& Wormleaton 2001).  For parameters with strong lensing (Figure
2) the transmission spectrum could be significantly affected by
atmospheric refraction. At wavebands corresponding to the
transparent continuum, strong lensing can lower the flux
(recall from \S \ref{explore} that lensing generally causes a net
suppression of flux during most of the transit), while as we
have discussed in the last subsection,
at wavebands corresponding to strong planetary absorption bands,
lensing has negligible effects. The contrast between continuum
and absorption lines is therefore decreased, and
effective line-strengths are therefore altered.

The magnitude of this effect can be estimated as follows.
Atmospheric lensing effect generally introduces a decrease in absolute flux
that is approximately $(R_0/R_*)^2 \times (H/R_0)$ for
planets with caustics.  Strong planetary atmosphere absorption lines
on the other hand cause a change in absolute flux that is approximately
$(R_0/R_*)^2 \times x \times (H/R_0)$, where $x \sim$ a few. Therefore,
in a planetary transmission spectrum, the effective line strength
is reduced from $x \times (H/R_0)$ to $\sim (x-1) \times (H/R_0)$.
The exact size of this effect depends on details of the atmospheric
structure, composition as well as the limb-darkening function.

As noted in our Figure~\ref{paramLENS}, and commented on previously
(Seager \& Sasselov 2000; Hubbard et al. 2001; Brown 2001) the
transmission spectra of the close-in extra-solar giant planets (EGPs)
---including the only
known transiting EGP HD~209458b---will be
little affected by refraction because caustics do not exist
due to the small planet-star distance (see \S\ref{explore}).

Finally, we end this section by briefly commenting on the
stellar spectrum.
During a planetary transit, the stellar spectrum changes with time, a
phenomenon referred to as a spectroscopic transit. There are at least
two different effects. The first has been discussed and observed by
Queloz et al. (2000). The planet blocks different
parts of the rotating stellar disk as the transit progresses, causing
red-shifting or blue-shifting of the stellar lines during the transit.
The second effect arises from the fact that different parts of spectral
lines and different lines are formed at different depths in the stellar
atmosphere. As the planet transits different parts of the limb-darkened
stellar disk, the stellar line shapes and strengths changes.
Atmospheric lensing introduces modifications to both of the above effects,
because flux from different parts of the star is magnified compared
to the nonlensed case.

\section{Discussion}
\label{discuss}

Our findings are summarized as follows.

\begin{itemize}
\item The importance of atmospheric lensing is mainly controlled by
two parameters: $R_0/H$ (the ratio of planet radius to atmospheric
scale height) and $B \equiv 2 \alpha \rho_0 D_{\rm LS} / H$
(product of the refractive coefficient, atmospheric density, and
star-planet separation divided by the scale height).\footnote{
In the solar system case, $D_{\rm LS}$ is replaced by $D_{\rm OL}$,
see \S \ref{solar}.} The condition for strong lensing---the existence of
caustics---is described by equation (\ref{strongcondition})
and depicted in Figure \ref{paramLENS}.

\item Strong lensing generally introduces a fractional change in the
light deficit of the order of $H/R_0$ during a planetary transit.  We
choose to discuss changes induced by lensing in terms of fractional
changes in the light deficit ($f(t)$ as defined in equation (\ref{ft}) and
shown in e.g. Figure \ref{dtransit22}) so that the absolute change in
observed flux can be easily estimated for any planet-to-star size
ratio.\footnote{The observed stellar flux here is always normalized by
its pre- or post-transit value.} It works as follows.  The drop in
flux during a nonlensed transit is about $(R_0/R_*)^2$ where $R_*$ is
the radius of the star. Strong lensing introduces an additional
absolute change in flux that is therefore about $(R_0/R_*)^2 \times
(H/R_0)$.

\item Lensing produces a characteristic slowing down in the dimming of
the star light at ingress (reversed at egress).  This is due to the
additional photons refracted into view from the planet's shadow.  The
light curve of a lensed transit can be mimicked to some extent by one
of a nonlensed transit, if parameters for the latter are appropriately
tuned.  We find, however, that the difference between the two can
remain significant especially during ingress and egress.  For example,
the dotted line in Figure \ref{dtransit22} shows a fractional
difference in the light deficit of $\sim 0.5 (H/R_0)$ at ingress or
egress. When this difference is larger than the observational
uncertainties, we should be able to constrain the lensing signal
parameters such as the atmospheric scale height, density and
refractive coefficient (parameters such as $R_0$ and $D_{\rm LS}$ can
be learned from other features of the light curve or other
observations).  The scale height and refractive coefficient will in
turn give us useful information on the temperature and chemical
composition of the planet.

\item The strength of lensing is expected to vary significantly with
color. The primary reason is the variation of absorption and
scattering with wavelength. In wavebands where absorption or scattering is
significant, caustics are effectively hidden by extinction.
Therefore, the optimal wavebands for detecting atmospheric lensing is
towards the red or near infrared, where the inevitable Rayleigh
scattering is less strong, but away from strong absorption bands such
as gaseous H$_2$O or CH$_4$.  Optically thick clouds (such as
enstatite in hotter planets or H$_2$O or CH$_4$ ice in cooler planets)
might contribute significant opacity. See Hubbard et al. (2001) for a
detailed treatment of HD~209458b.

\item Several conditions are therefore advantageous for detecting the
lensing signal: 1) observations at wavebands where absorption and
scattering is weak; 2) a high temperature of the planet atmosphere due
to a nearby hot star, which raises the atmospheric scale height
(thereby increasing $H/R_0$); 3) $D_{\rm LS}$ large enough to keep $B
\equiv 2 \alpha \rho_0 D_{\rm LS}/H$ above the strong-lensing
threshold.  Our fiducial model (solid square in Figure
\ref{paramLENS}) provides an example where lensing induces a
fractional change in the light deficit of around $1\%$. For a planet
that has an area about $1\%$ of the star, this implies lensing causes
an absolute change of about $10^{-4}$ in the normalized flux---an
observationally relevant effect. This model can be realized by a
gaseous giant (similar to Jupiter in mass, but with a slightly
inflated radius due to high temperature) about an AU away from an A
star, with observations done in red wavebands ($\lambda \sim 10^4$~\AA), but away from possible strong absorption bands such as
water. We should emphasize there is considerable uncertainty in the
size of such a planet, because the radius of a hot gaseous giant
varies significantly with time (Burrows et al. 2000; see also Murray
et al. 1998 and Lin et al. 1998 on change in orbital characteristics
over time).

\item Oblateness of the planet induces an asymmetry to the transit
light curve (about the point of minimum flux), which vanishes only
when the semi-major or semi-minor axis of the planet is exactly
aligned with the direction of relative motion. The asymmetry is
about $0.5 \%$, as measured in fractional light deficit (equation
(\ref{fasym}) and Figure (\ref{asymm})), for a projected oblateness or
ellipticity of
$\epsilon = 0.1$, and an angle of $45^0$ between the semi-major axis
and direction of motion.  The absolute difference in flux between
ingress and egress is therefore about $0.5 \% \times (R_0 / R_*)^2$,
and is for example $5 \times 10^{-4}$ for a planet that is a tenth the
size of the star.

\item For HD 209458b, the only currently known transiting extrasolar
planet, caustics and therefore strong lensing is absent, 
because of the very small lens-source separation (and the resulting
large scale height due to a high temperature). 
Its oblateness is expected to be small, $\lsim 10^{-3}$, due
to the tidal locking of its rotation to orbital motion.

\end{itemize}

In this paper, we have focused on the effects of atmospheric lensing
and planet oblateness on the transit light curve. It is worthwhile to
briefly list other possible ``secondary'' fluctuations to the light
curve---variations in the observed flux other than those due to the
standard spherical occultation of a star with a smooth limb darkening
profile. They can be divided into three categories.

The first category is due to close companions of the planet.
Rings and moons, if present,
induce an asymmetry to the light curve that may be confused
with oblateness. They also introduce additional parameters
that control the light curve, making it perhaps more difficult to
discern the lensing signal. Taking Jupiter as an example,
the largest moon Ganymede has a radius of $2635$ km, about
$3.7 \%$ the size of Jupiter. The same moon of
a Jupiter-like planet orbiting another star
would induce a fractional change of about $10^{-3}$
in the light deficit. This is about $5 - 10$ times smaller
than the lensing or oblateness effects we found for our most
optimistic models (Figures \ref{dtransit22} and \ref{asymm}).
If the moon is sufficiently far from the planet, it might
also introduce distinct signatures at ingress or egress
(e.g. Sartoretti \& Schneider, 1998) that
can be disentangled from oblateness or lensing effects.
An opaque ring 1/2 the size of Saturn's rings
for a gaseous giant with $R_p$ = 1.4 $R_J$ orbiting a sun-like star
would deepen the planet transit light curve by 0.05\% - 1.2\%
depending on the inclination of the ring
(Seager \& Hui 2000).
Ring effects might be isolated by looking for
distinct shapes in the transit light curve (see Brown et al. 2001),
and possibly rescattered star light.

The second category is due to the planet itself. The planet atmosphere
can have persistent or transient disturbances that modify the transit
light curve. Such disturbances have to be large in extent to change
the light curve significantly. For instance, small uniformly
distributed clouds would not introduce an asymmetry to the light
curve.  Something like the Great Red Spot on Jupiter could conceivably
be large enough. But such disturbances cannot introduce a fractional
change in the light deficit that is larger than the ratio of the
projected atmospheric area to total planetary area i.e. $\sim H/R_0$.
It is important to emphasize that perturbations to the atmosphere that
only change the effective radius of the planet $R_0$ are not
sufficient to wash out the lensing signal; as we have shown in \S
\ref{signature}, just tuning $R_0$ is not enough to confuse a lensed
transit with a nonlensed transit that has different
parameters. Hubbard et al. (2001) pointed out that Rayleigh
scattering, in addition to causing extinction, also produces a glow
around the planet, which changes its effective size (fractional
variation in size about $1\%$ between different wavelengths). It would
be interesting to explore how both Rayleigh and condensate
rescattered stellar
photons affect the detailed shape of the transit light curve
(i.e. aside from a simple change in $R_0$).  Lastly, we have also
ignored diffraction in this paper. Diffraction is likely unimportant
for most extrasolar cases of interest, for two reasons.  First, the
relevant parameter combination $(R/H)^{1/2} B D_{\rm OL}/D_{\rm LS}$
is greater than unity (see Fig. \ref{paramLENS}), and is therefore not
favorable for diffraction (see e.g. French \& Gierasch 1976, Elliot et
al. 1975). Second, the observed flux comes from a sum over incoherent
sources distributed over the stellar surface.

The third category of secondary fluctuations is due to complications
in the star.  Stellar oscillations are expected to produce absolute
changes in the flux at a $10^{-5}$ level on the time-scale of a
transit (see e.g.  website for the proposed Kepler mission {\tt
http://www.kepler.arc.nasa.gov/}).  Also, realistic stellar profiles
might have bumps and wiggles on top of the smooth profile we have
assumed (\S \ref{limpdark}). For instance, the light curve would be
modified if a planet happens to transit over a star spot (e.g. Seager \& Hui
2000).
The largest effect is obtained if the spot has a size that
is comparable to the planet, and the planet happens to overlap
completely with the spot during the transit. At the point where the
planet and spot coincide, the flux would basically return to unity,
its pre-transit value. For example, an Earth-sized planet which
happens to cross over an Earth-sized star spot would produce
this behavior, as illustrated in the upper panel of Figure \ref{spot}.
This kind of situation is probably rare, however,
particularly if the planet considered is a gaseous giant i.e. much
larger than typical star spot (lower panel of Figure \ref{spot}).
Then, the absolute change in flux in
such coincident transits will be roughly the ratio of the spot area to
star area, $\lsim 10^{-5}$ borrowing from the example of the Sun.
Note also that while lensing or oblateness effects are most easily
recognized at ingress or egress, changes to the light curve due to
imperfections in the stellar profile can occur throughout the transit.
More detailed studies would be required to see if spots produce
different signatures from lensing or oblateness, if the planet
crosses them  at ingress or egress. Note also
that effects due to star spots are likely not repeatable
over many planetary periods.

It is also worth emphasizing that even in cases where the lensing or
oblateness signal is weak and therefore not easily isolated, it
could still act as an important contaminant in confusing other signals
one might be interested in, such as detection of moons and planetary
rings. From the above discussions,
it is clear future high precision measurements of extrasolar planetary
transits will present a very interesting challenge---there are several
sources of small secondary fluctuations (change in flux of $10^{-4}$
or less) which are within observational reach and which would require
some effort to disentangle. The rewards of such efforts will be
substantial---from detection of moons and rings, to measuring the
(projected) planet oblateness and therefore constraining its
rotational period; from studies of the stellar atmosphere to studies
of the planetary atmosphere, constraining for instance its scale
height, temperature, density and composition.

\acknowledgements{We are grateful to John Bahcall, Bruce
Draine, Scott Gaudi, Joe Patterson, and
Penny Sackett for useful discussions.  SS is
supported by the W. M. Keck foundation, and LH is supported by the
Taplin Fellowship at the IAS and an Outstanding Junior Investigator
Award from the DOE.}

\section*{Appendix A}
\label{appendixA}

Here, we would like to motivate the density profile we adopt in this
paper.  We begin by the following statement of hydrostatic equilibrium
of the atmosphere:
\begin{equation}
\nablabf P = -\rho \nablabf \phi_{\rm eff.}
\end{equation}
where $P$ is the pressure, and $\phi_{\rm eff.}$ is an effective potential
which is the sum of the gravitational potential $\phi$ and
a potential to take into account rotation:
\begin{equation}
\phi_{\rm eff.} = \phi - {1\over 2} \omega^2 ({x'}^2 + {y'}^2)
\end{equation}
where $z'$ is the axis of rotation, $x'$ and $y'$ are
the perpendicular axes (we will use $x'_i$, $i=3,1,2$ to
denote them) and $\omega$ is the angular speed.
For an isothermal atmosphere where $P \propto \rho$, the above implies
\begin{equation}
\rho \propto {\rm exp} [-\phi_{\rm eff.}]
\label{rhophi}
\end{equation}
We note that planetary atmospheres are only approximately isothermal.
Deviations such as temperature inversion are known to exist (e.g. Jones
1999).

The question then reduces to what kind of gravitational potential one
expects for a
general rotating figure of equilibrium. For bodies that
are not too aspherical, we can expand the gravitational potential
using the Legendre polynomials (see e.g. Danby 1962, Chandrasekhar 1969)
\begin{equation}
\phi = - {G M_p \over r'} \left( 1 - {J_2 \over {r'}^2} {1\over 2}
(3 {\rm cos}^2 \theta - 1)
- ... \right)
\end{equation}
where $M_p$ is the planet mass, $J_2$ is some constant coefficient which is
presumably
small, $\theta$ is the angle
between the radial vector and the $z'$ axis, and $r' = \sqrt{
{x'}^2 + {y'}^2 + {z'}^2}$. It is sufficient to illustrate
our argument using only the first term -- it is possible to
generalize to include higher order terms in
the potential expansion.

We are interested in the form of $\phi_{\rm eff.}$ when
$r' = R (1+\eta)$ with $\eta \ll 1$, where $R$ is
approximately the planet radius. We are also interested in
cases where the angular speed $\omega$ is in some sense small
--- the relevant parameter to consider is $R^3 \omega^2 / (2 G M_p)
\equiv \delta \ll 1$.
One can then express $\phi_{\rm eff.}$ as
\begin{eqnarray}
\phi_{\rm eff.} &=& - {G M_p \over r'} - {1\over 2} \omega^2
({x'}^2 + {y'}^2) \\ \nonumber
&\sim& - 2 {G M_p \over R}
+ {G M_p \over R^2} \sqrt{{x'}^2 + {y'}^2 + {z'}^2} - {1\over 2} \omega^2
({x'}^2 + {y'}^2) \\ \nonumber
&\sim& - 2 {G M_p \over R}
+ {G M_p \over R^2} \left[ {{x'}^2 \over (1 + \delta)^2} +
{{y'}^2 \over (1 +\delta)^2} + {{z'}^2 }\right]^{1/2}
\end{eqnarray}
which is exact up to terms of first order
in $\eta$ and $\delta$. Clearly, $\phi_{\rm eff.}$ is
a function of $r$ alone, where $r^2 \equiv \sum_i {x'_i}^2 / {a_i}^2$,
where $a_1 = a_2 = 1+\delta$, $a_3 = 1$.
In the above derivation $a_1 = a_2 \ne a_3$.
A rotating body of self-gravitating fluid
can actually be triaxial in general, but
triaxiality is likely unimportant for the rotational
velocities of interest here (see Chandrashekhar 1969, Bertotti \&
Fasinella 1990).

Putting the above into equation (\ref{rhophi}), we obtain
\begin{equation}
\rho = \rho_0 {\,\rm exp} [-(r-R_0)/H]
\end{equation}
where $H^{-1}$ is the derivative of the effective potential,
and $\rho_0$ is the atmospheric density at $r = R_0$. The choice of $R_0$ is
arbitrary at this point. In \S \ref{formalism}, we choose it to be the
radius
below which the atmosphere is completely opaque.

Finally, note that the principle axes defined by $x'_i$ are not necessarily
lined up with
the axes defined by $x_i$ in \S \ref{formalism} by the lensing geometry.
The two sets are generally related by a rotation.
Since there is freedom in rotating $x_1$ and $x_2$ (i.e. the lensing
geometry only picks out $x_3$ or the z direction), we can without loss of
generality
relate the two sets of coordinates by two rotation matrices:
${\bf x'} = {\bf R_{z'}} (\gamma) \cdot {\bf R_{x'}} (\beta) \cdot {\bf x}$,
where ${\bf R_{z'}} (\gamma)$ is a rotation about $x'_3$ by angle $\gamma$,
and
${\bf R_{x'}} (\beta)$ is a rotation about $x'_1$ by angle $\beta$.
Using $a_1 = a_3 (1 + \epsilon_A)$ and $a_2 = a_3 (1 + \epsilon_B)$, and
assuming $\epsilon_A$ and $\epsilon_B$ are small (in our
above derivation, $\epsilon_A = \epsilon_B = \delta$ i.e. an oblate
spheroid; our derivation below continues to work even if
this were violated), we can write
\begin{eqnarray}
\label{rappendix}
r^2 = \tilde z^2 + (1 - 2 \epsilon_A {\rm cos}^2 \gamma - 2 \epsilon_B {\rm
sin}^2 \gamma) x^2
+ 4 (\epsilon_A - \epsilon_B) x y +
(1 - 2 \epsilon_A {\rm sin}^2 \gamma {\rm cos}^2 \beta - 2 \epsilon_B {\rm
cos}^2 \gamma
{\rm cos}^2 \beta) y^2
\end{eqnarray}
where $\tilde z = [1 + O(\epsilon_A,\epsilon_B)] [z +
O(\epsilon_A,\epsilon_B) x
+ O(\epsilon_A,\epsilon_B) y]$.

Now, recall that the quantity that we are interested in, the deflection
angle $\theta_D$,
is given by equation~(\ref{thetaD}): $\theta_D = \nablaperp
\int_{-\infty}^\infty
 \alpha \rho(r) dz$. One can clearly change the variable of integration from
$z$
to $\tilde z$: $\int_{-\infty}^\infty
 \alpha \rho_0 {\,\rm exp} [-(r-R_0)/H] dz = \int_{-\infty}^\infty
 \alpha \rho_0 {\,\rm exp} [-(r-R_0)/H] d\tilde z$ where we have absorbed
the slight change in multiplicative factor into a redefinition of $\rho_0$.
Moreover, since $\tilde z$ is a dummy integration variable, we could as well
rename $\tilde z \rightarrow z$ in equation~(\ref{rappendix}).

Consider next the terms involving $x$ and $y$ in equation~(\ref{rappendix}).
We could easily
perform a rotation to put them in the simple form: $(1 - \epsilon) x^2 + (1
+ \epsilon) y^2$
if we absorb multiplicative factors into a redefinition of $H$ (such
multiplicative
factors would affect $z$ also, but they can once again be absorbed into
redefinition of
$\rho_0$).
To be more specific, let us consider the important case where $\epsilon_A =
\epsilon_B$.
Then, the
$x$ and $y$ terms reduce to
\begin{eqnarray}
(1 - 2 \epsilon_A) x^2 +
(1 - 2 \epsilon_A {\rm cos}^2 \beta) y^2
= [1 - \epsilon_A (1 + {\rm cos}^2 \beta)]
[ (1-\epsilon) x^2 + (1+\epsilon) y^2 ]
\end{eqnarray}
where
\begin{eqnarray}
\epsilon = \epsilon_A (1 - {\rm cos}^2 \beta)
\end{eqnarray}
For a random distribution of angle $\beta$, we expect on the average
$\epsilon = \epsilon_A / 2$. It is unclear, however, if $\beta$ should
be randomly distributed. The solar system planets do seem to have
rotational axes pointing in all kinds of directions with respect to
their orbital planes. The quantity $\epsilon_A$ describes directly
the shape of the planet: contours of constant density obey
${z'}^2 + ({x'}^2 + {y'}^2)/(1+ \epsilon_A)^2 =$ constant. In other words,
oblateness,
defined by the ratio $(a_1 - a_3) / (a_1 + a_3)$, is given by
$\epsilon_A/2$ to the lowest order. One can view $\epsilon$ as a kind of
projected
ellipticity or oblateness.

To summarize, with suitable rescaling of $z$, $H$ and $\rho_0$,
we have
\begin{eqnarray}
\theta_D = \int_{-\infty}^\infty \alpha \rho_0 {\,\rm exp} [-(r-R_0)/H] dz
\quad , \quad r^2 = (1 - \epsilon) b_1^2 + (1 + \epsilon) b_2^2 + z^2
\end{eqnarray}
where we have equated $x$ and $y$ with impact parameters $b_1$ and $b_2$ as
defined in \S \ref{formalism}.

\section*{Appendix B}
\label{appendixB}

We discuss here how to solve equation~(\ref{lensmaster}) and how to
find the caustic and critical curve. Equation~(\ref{lensmaster}) has a
form that is exactly analogous to elliptic potentials sometimes used in
modeling gravitational lenses.  The main trick for solving this type
of lensing equation is taken directly from Schneider, Falco \& Ehlers
(1992), but we provide additional comments here for cases that require
special attention.

Equation~(\ref{lensmaster}) is a set of two equations for $\theta_I^1$ and
$\theta_I^2$, given
the source position $\thetaS$.
Eliminating $\psi(u)$ from the two equations, one obtains
\begin{equation}
\theta_I^2 = {(1-\epsilon)\theta_I^1 \theta_S^2 \over (1+\epsilon)\theta_S^1
- 2\epsilon\theta_I^1}
\label{thetaI2}
\end{equation}
This gives us $\theta_I^2$ as a function of $\theta_I^1$, and substituting
into
the first component of the lensing equation leaves us with a single equation
for $\theta_I^1$:
\begin{equation}
\theta_S^1 = \theta_I^1 + (1-\epsilon) \theta_I^1 \psi(u)
\label{thetaS1}
\end{equation}
where $u$ is a function of $\theta_I^1$ and $\theta_I^2$ (now function of
$\theta_I^1$)
as well. The lensing problem is therefore no more complicated than one with
spherical symmetry. Once $\theta_I^1$ is solved for, $\theta_I^2$ can be
obtained from
equation~(\ref{thetaI2}).

For the case of $\theta_S^2 = 0$, the above solution always puts $\theta_I^2
= 0$.
There is, however, another possibility when multiple images are allowed.
Examining the analog of equation~(\ref{thetaS1}) for the second component,
one can
see that the other option for a vanishing $\theta_S^2$ is to have
\begin{equation}
1 + (1+\epsilon) \psi(u) = 0
\end{equation}
This gives a single equation for $\theta_I^2$ if one uses the fact that
$\theta_I^1 = (1+\epsilon) \theta_S^1 (2\epsilon)^{-1}$. This is obtained
from eliminating $\psi(u)$ from the two components of
equation~(\ref{lensmaster}), but
noting $\theta_S^2 = 0$ and assuming $\theta_I^2 \ne 0$.

There is another special case: $\theta_S^1 = 0$. In this case,
equation~(\ref{thetaS1})
tells us that there are two possibilities: $1 + (1-\epsilon) \psi(u) = 0$ or
$\theta_I^1 = 0$. The former is handled automatically with the computational
procedure above. The latter requires more care: one can solve
\begin{equation}
\theta_S^2 = \theta_I^2 + (1+\epsilon) \theta_I^2 \psi(u)
\end{equation}
for $\theta_I^2$ by setting $\theta_I^1 = 0$. Doing so ensures
all possible images are uncovered.

To find the caustic and critical curve, we solve $A^{-1} = 0$ using
equation~(\ref{A}):
\begin{eqnarray}
\label{causticcond}
v = {-1 \over \epsilon \tilde\psi} [1 + 2\psi + (1-\epsilon^2)\psi^2 +
(1-\epsilon^2)u^2
\psi\tilde\psi + u^2 \tilde\psi]
\end{eqnarray}
This gives us $v$ for a given $u$. However, from the definition of $v$ and
$u$ in
equation~(\ref{A}), it is clear only $|v| \le u^2$ is physical. Therefore,
the caustic or
critical curve corresponds to solutions of $v$ given $u$ that satisfies $|v|
\le u^2$.
Strong lensing occurs when such solutions exist.
For each pair of $v$ and $u$, it is easy to solve for $\theta_I^i$ from
their definitions:
\begin{eqnarray}
\theta_I^1 = \pm \sqrt{{u^2 - v \over 2 (1-\epsilon)}} \quad , \quad
\theta_I^2 = \pm \sqrt{{u^2 + v \over 2 (1+\epsilon)}}
\end{eqnarray}
Given the critical curve defined by $\theta_I^i$, one can then solve for the
caustic
using the lens mapping (equation~(\ref{lensmaster})). We note that
the above two expressions were interchanged by mistake in Schneider et al.
(1992).

\begin{table}
\begin{center}
\begin{tabular}{|ccc|}\hline
Gas comp. & $A_1 \times 10^5 {\rm cm^3}$ & $B_1 \times 10^{11} {\rm
cm^{-2}}$ \\ \hline \hline
Hydrogen & $13.6$ & $7.7$ \\
Oxygen & $26.63$ & $5.07$ \\
Nitrogen & $29.19$ & $7.7$ \\
Air & $28.79$ & $5.67$ \\
Methane & $42.6$ & $14.41$ \\ \hline
\end{tabular}
\end{center}
\caption{\label{alphatab} Refractive coefficients for different gas
compositions.
Refractive index is $n = 1 + \alpha \rho$, with $\alpha \rho = A_1 (1 +
B_1/\lambda^2)$.
Values are given for gases at $15^0$ C and $1$ atm. unit of pressure.}
\end{table}

\newpage
\begin{figure}
\plotone{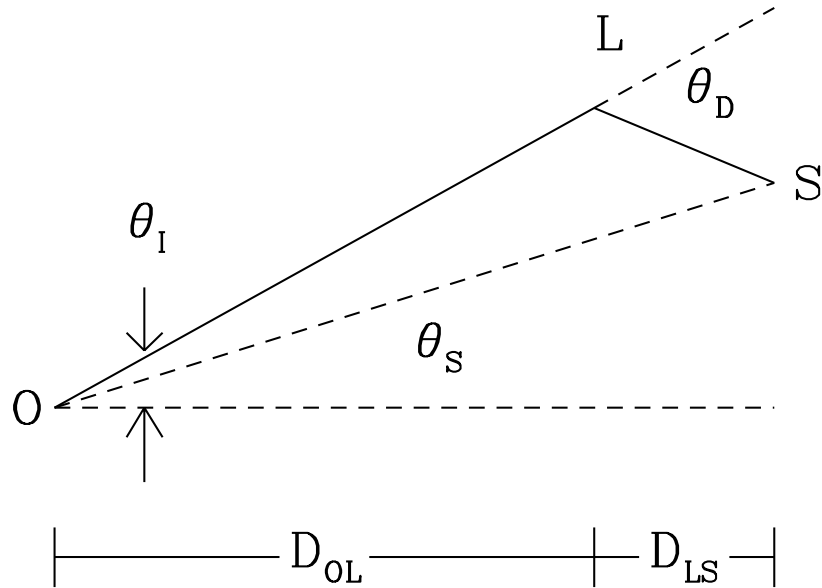}
\caption{A schematic ray-tracing diagram:
the solid line joining O (observer), L (lens)
and S (source) represents the light ray.
L represents the point of closest approach in
the transiting planet's atmosphere. S represents a point on
the surface of the star. The deflection is
exaggerated and the diagram is not to scale.
We use the convention that $\thetaD$ has an opposite sign
to $\thetaI$ and $\thetaS$.}
\label{fig:raytrace}
\end{figure}

\begin{figure}
\centerline{\psfig{figure=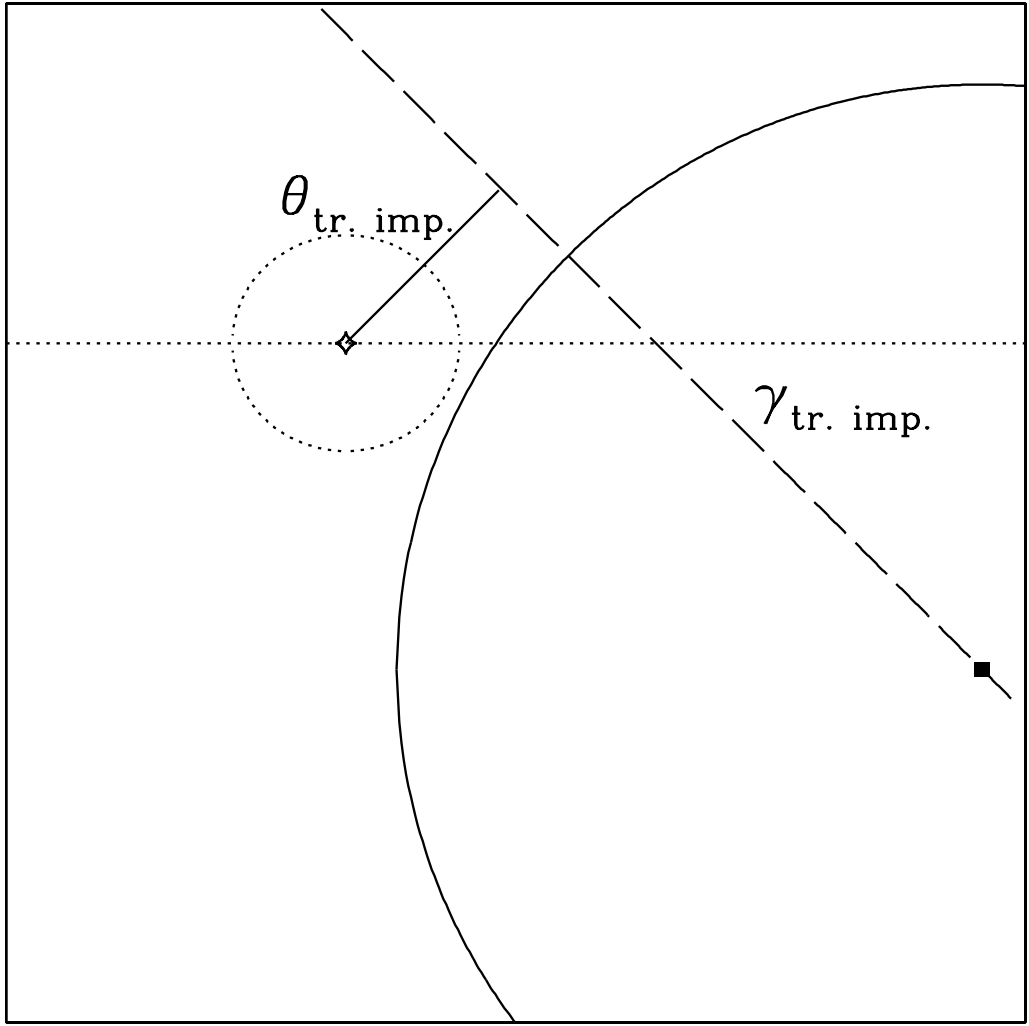,height=7.0in}}
\caption{\label{geometry} Geometrical set up of a planetary transit.
The circular solid line represents the boundary of the star, while the
ellipsoidal dotted line represents that of the planet.  The solid
square is the center of the star, which moves along the dashed line
with respect to the planet. The horizontal dotted line defines the
major axis of the planet. $\gammaimp$ is the angle between the
horizontal and dashed line, while $\thetaimp$ is the distance in
angular units of closest approach between the planet center and the
stellar center.}
\end{figure}

\begin{figure}
\centerline{\psfig{figure=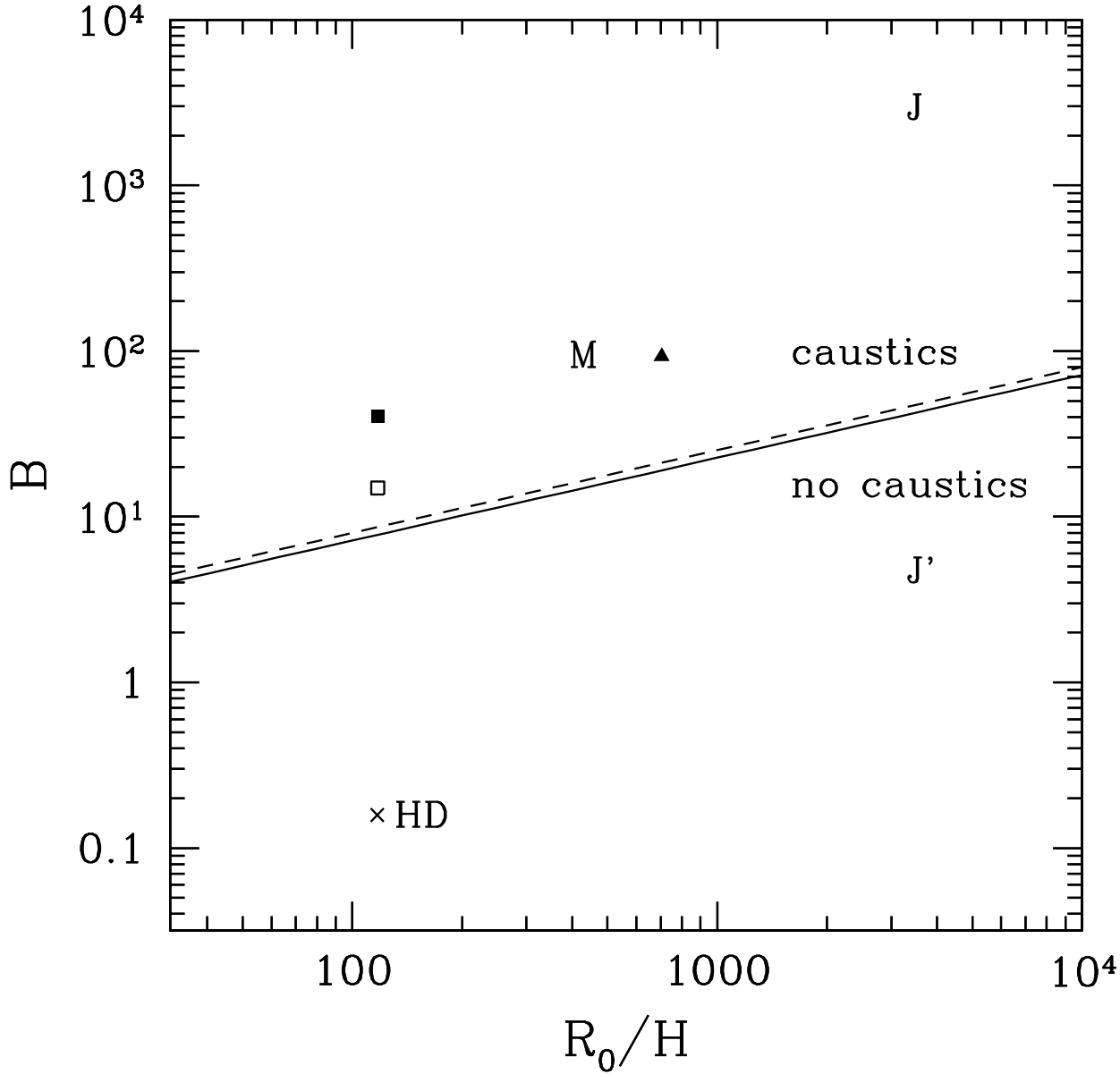,height=7.0in}}
\caption{\label{paramLENS} The solid and dashed lines delineate the
separation of the strong lensing regime (existence of caustics) from
the weak lensing regime, for the combination of parameters $B$
and $R_0/H$. The solid line is for projected oblateness/ellipticity
$\epsilon =
0.1$ and the dashed line is for a sphere, $\epsilon = 0$.  The
parameter $B$ is defined as $B = 2 \alpha (\rho_0/H) D_{\rm LS} D_{\rm
OL} / (D_{\rm OL} + D_{\rm LS})$ (equation~(\ref{psirewrite})).  The
quantity $R_0/H$ is the ratio of planet radius to atmospheric scale
height.  We show values of $B$ and $R_0/H$ for HD~209458b (HD), where
$D_{\rm LS} \ll D_{\rm OL}$, and for Mars (M) and Jupiter (J), where
$D_{\rm LS} \gg D_{\rm OL}$. In all cases, observations at
optical wavebands are assumed. We show with the symbol $J'$ the parameters
for Jupiter if it were observed at wavebands with 600 times stronger
absorption (e.g. in the UV, see text for details).
The solid square denotes the fiducial model we study in this paper.
The triangle, open square and cross are other models we also discuss.}
\end{figure}

\begin{figure}
\centerline{\psfig{figure=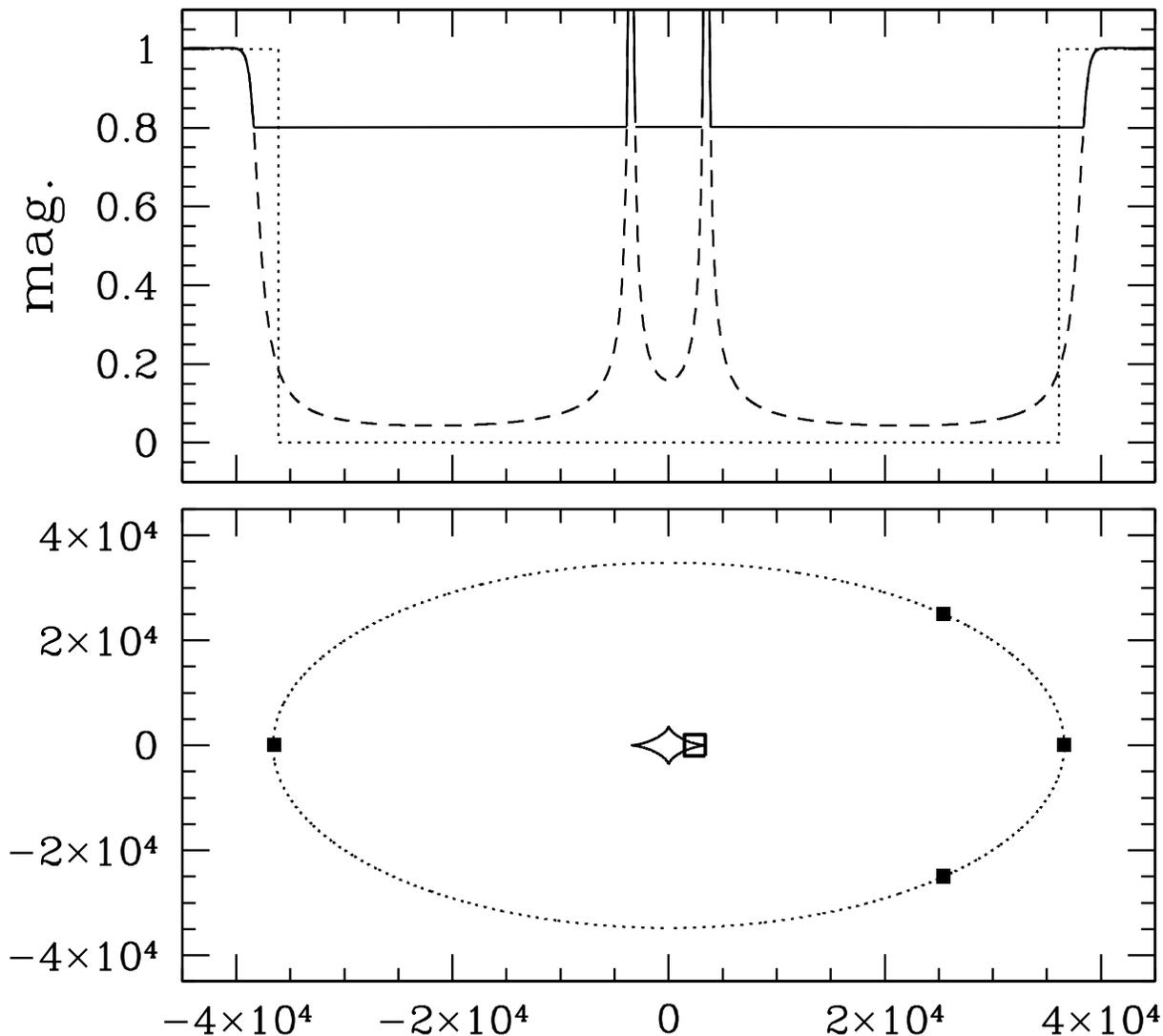,height=7.0in}}
\caption{\label{caustic2}
Upper Panel: The dashed line (overlaps with solid line
in the upper portion) shows the
magnification as a function of source position, for source positions
that lie on the major axis of the planet (the x-axis in the lower
panel.) The dotted line shows the occultation kernel
$W$ as defined in equation~(\ref{W}). The solid line represents
the light curve for an occultation by a solar system planet
of a distant star, which happens to be moving along the planet's
semimajor axis. See text for details.
Lower Panel: The solid line shows the caustic,
source positions where the magnification diverges, while the dotted line
shows
the critical curve, which is the image of the caustic. The four solid
squares are images of a point source denoted by an open square
situated just inside the caustic. The axis coordinates are distances
(in km) in the lens plane, i.e. they correspond to physical distances
from the planet center.}
\end{figure}

\begin{figure}
\centerline{\psfig{figure=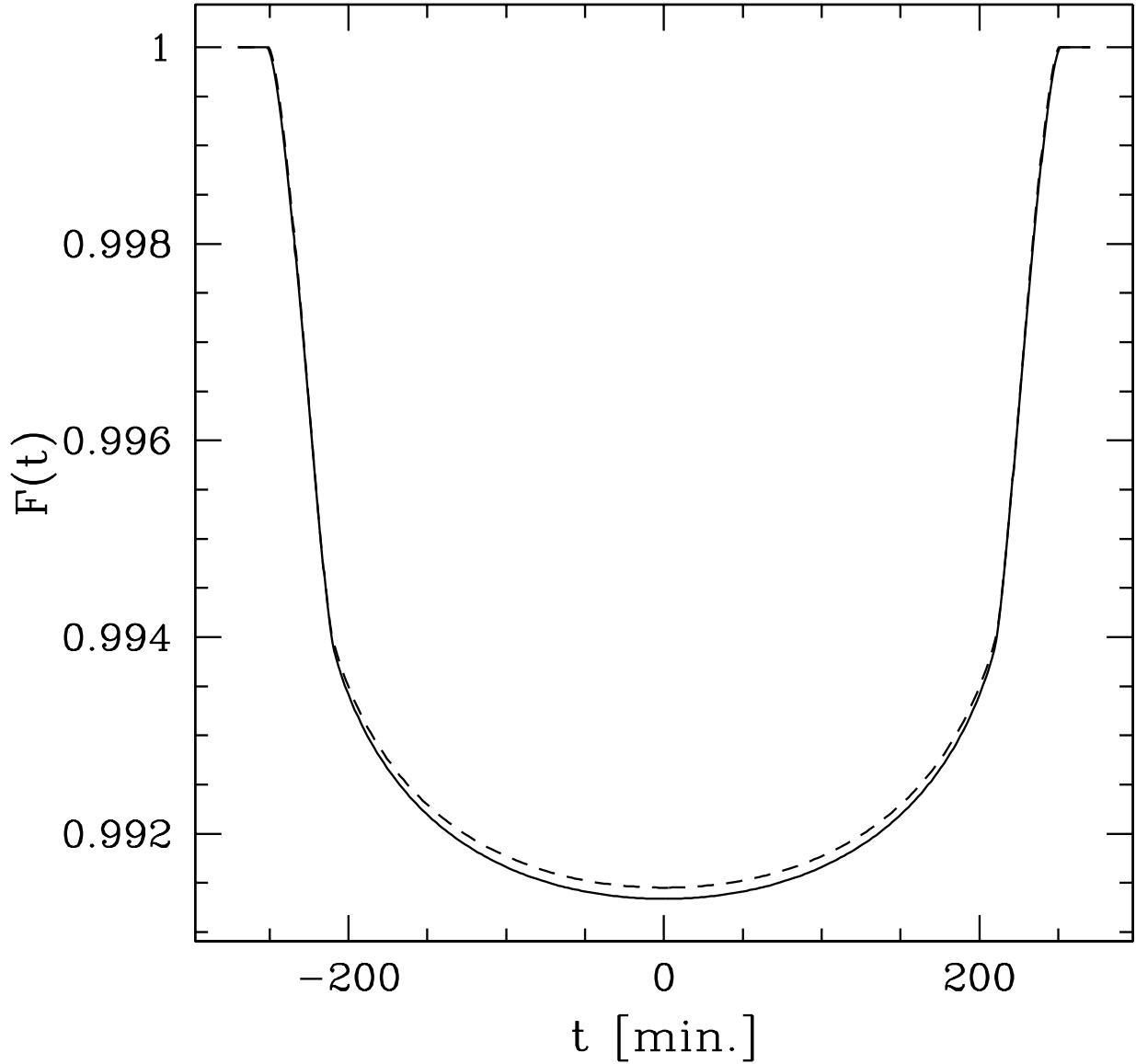,height=7.0in}}
\caption{\label{transit22} A transit light curve for our fiducial
model (denoted by a solid square in Figure~\ref{paramLENS} with
atmospheric lensing (solid curve) and without (dashed curve).  F(t) is
the normalized stellar flux (equation~(\ref{F})).
The star here has $R_* = 0.6 R_\odot$.
}
\end{figure}

\begin{figure}
\centerline{\psfig{figure=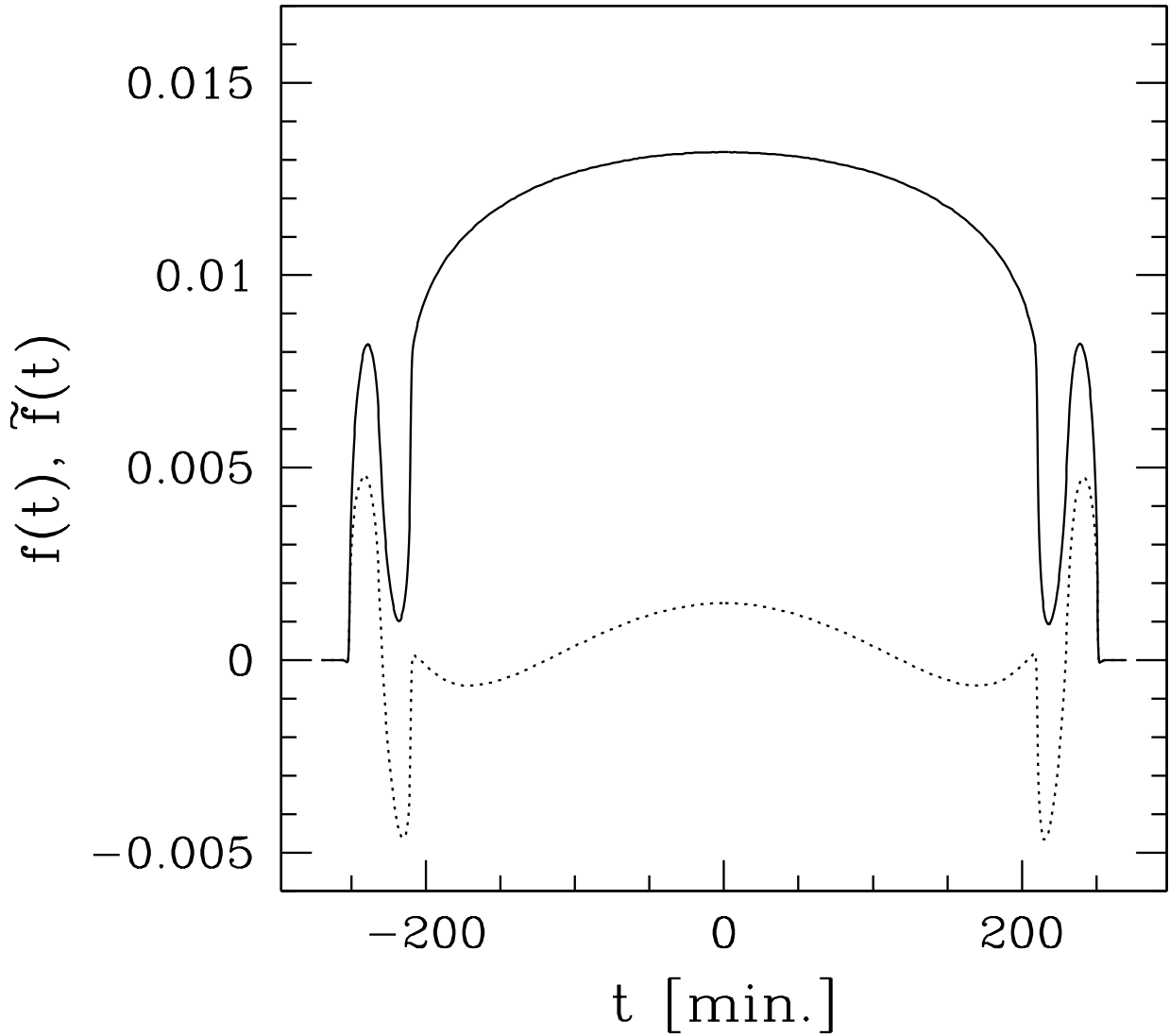,height=7.0in}}
\caption{\label{dtransit22} The solid line shows $f(t)$, the
fractional difference between the flux deficit from
a stellar light curve with and without planetary atmospheric
lensing (equation~(\ref{ft})). This is for the fiducial model denoted
by the solid square in Figure \ref{paramLENS}. The dotted line shows
$\tilde f(t)$ as defined in equation~(\ref{ftilde}), the fractional
difference between the light deficit from a
lensed fiducial model and an unlensed model
with parameters of the unlensed model tuned to minimize this
fractional difference: $R_0 = 35400$ km, $H = 300$ km, $R_* = 417600$
km, $u_* = 0.832$, $v_* = -0.2475$, $\rho_0 = 1.68 \times 10^{-5}$,
$\epsilon = 0.05$, $w = 2.9741 \times 10^{-13} {\,\rm s^{-1}}$,
$\gammaimp = 49.5^0$, and $\thetaimp = 69696 {\,\rm km} / D_{\rm OL}$.
}
\end{figure}

\begin{figure}
\centerline{\psfig{figure=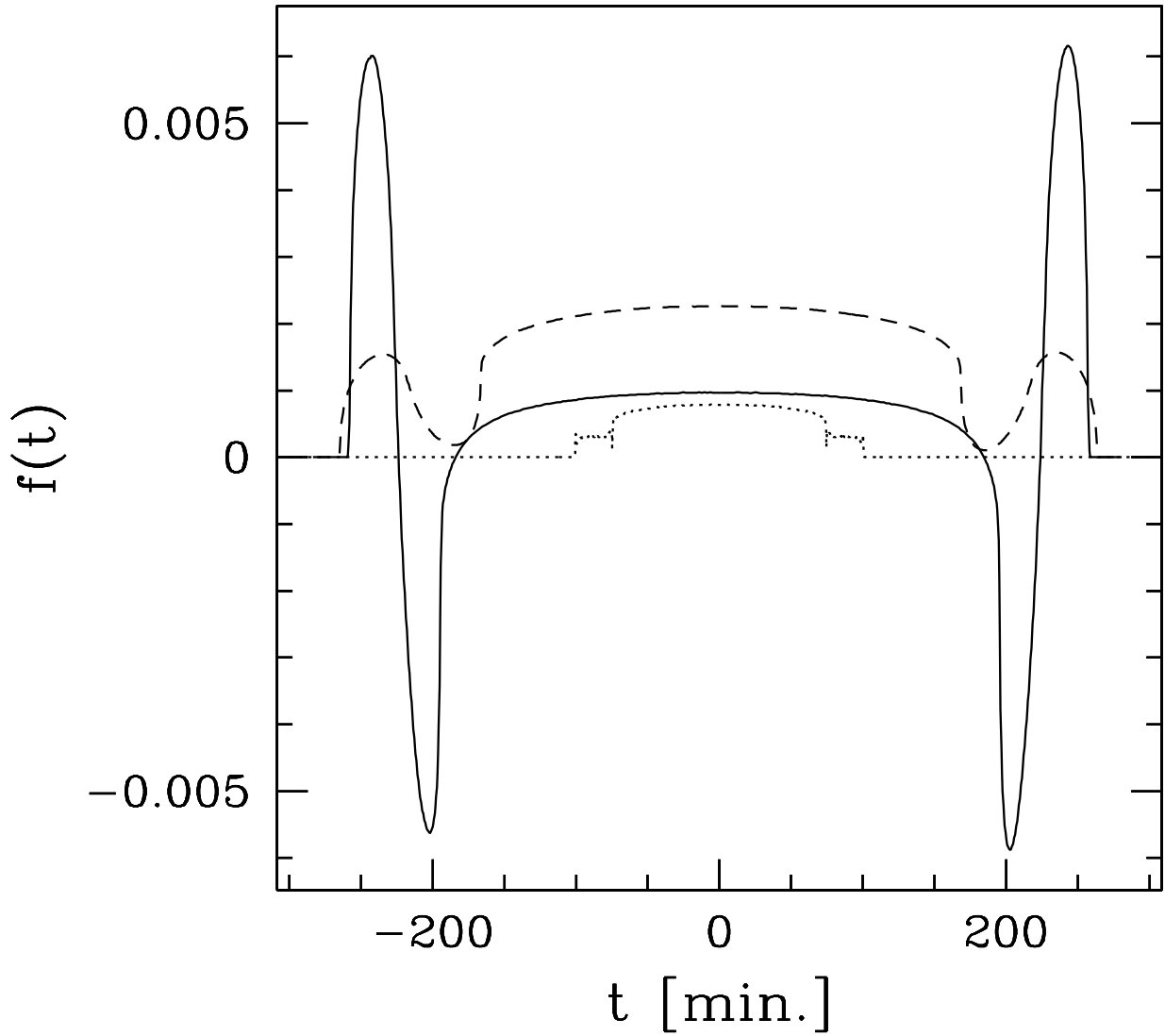,height=7.0in}}
\caption{\label{dtransit22.5}
$f(t)$ (equation~(\ref{ft})) for three different
planetary models corresponding to the
open square (solid line), cross (dotted line), and triangle
(dashed line) in Figure \ref{paramLENS}. Parameters
for the dotted line resemble those for HD 209458b.}
\end{figure}

\begin{figure}
\centerline{\psfig{figure=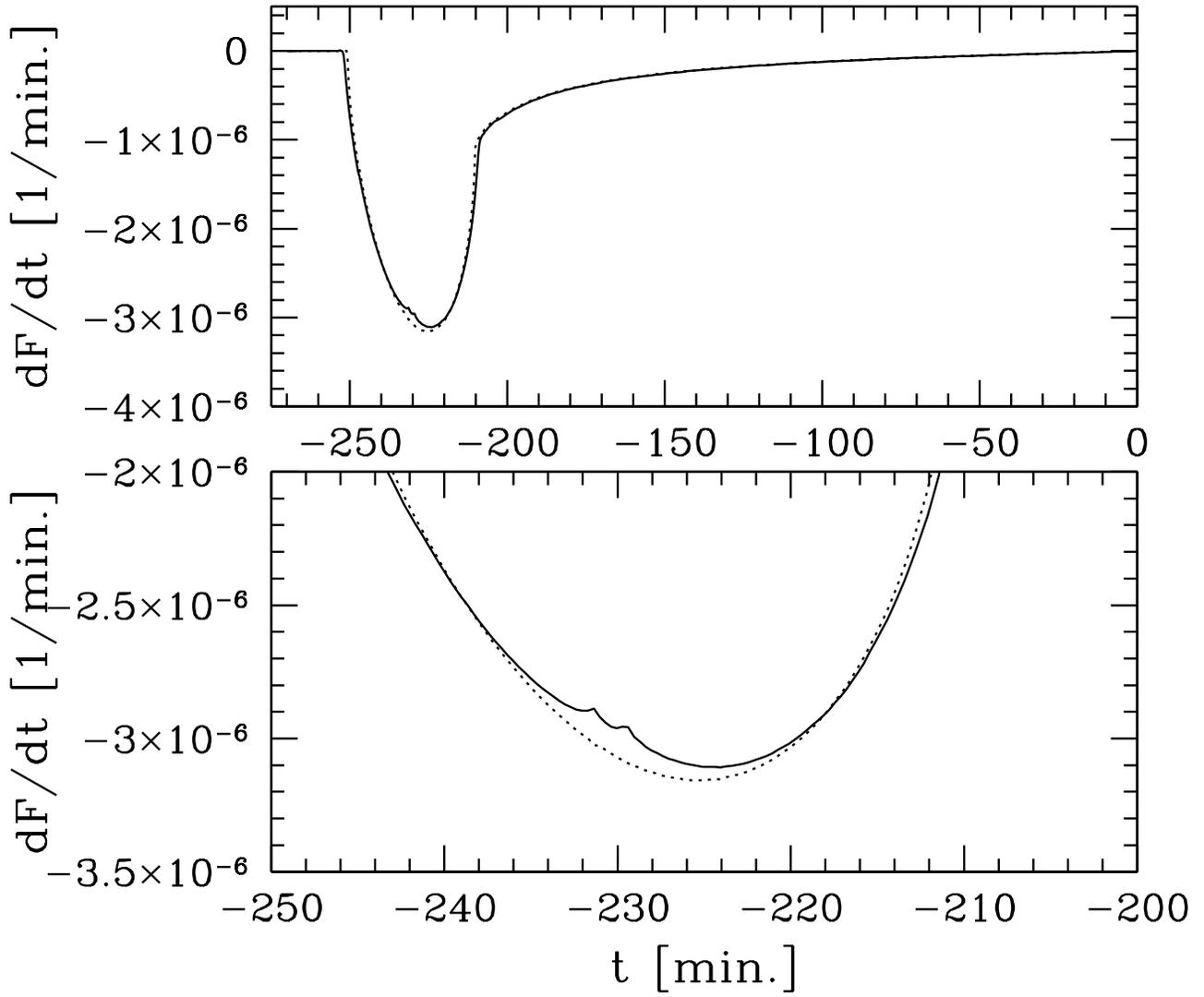,height=7.0in}}
\caption{\label{transitDnew22} $dF/dt$ for the fiducial model with $\epsilon
= 0.05$, denoted
by the solid square in Figure \ref{paramLENS}. The lower panel
is a magnified version of the upper one. The solid line
is with lensing, and the dotted line is without.
}
\end{figure}

\begin{figure}
\centerline{\psfig{figure=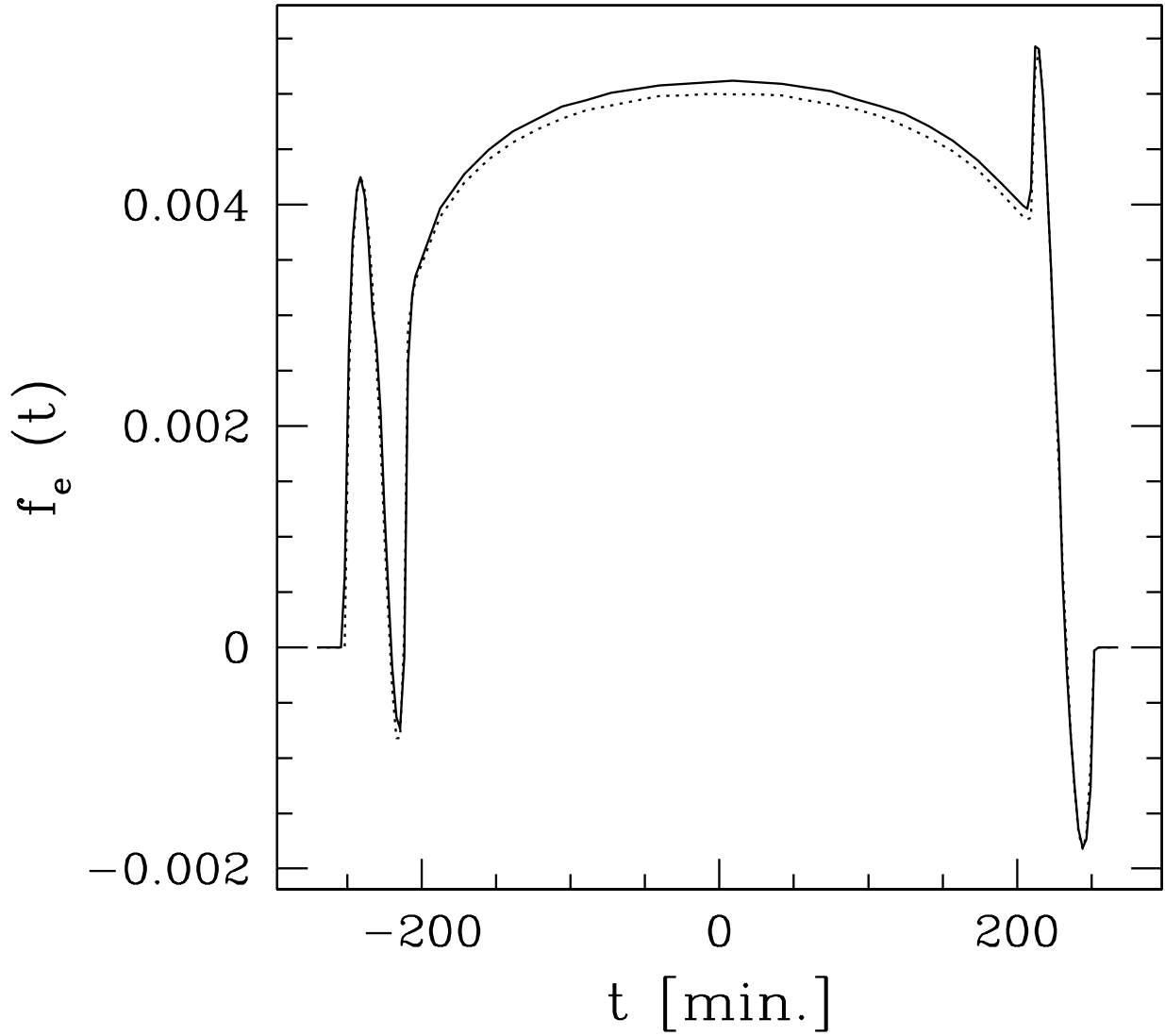,height=7.0in}}
\caption{\label{diff33} The two curves show $f_e$
(equation~(\ref{fe})), the fractional difference between the light
deficit from a transit of an elliptical planet and a spherical planet,
for the fiducial model denoted by the solid square in Figure
\ref{paramLENS}. The solid line is for a transit with lensing, and
dotted line for a transit without.}
\end{figure}

\begin{figure}
\centerline{\psfig{figure=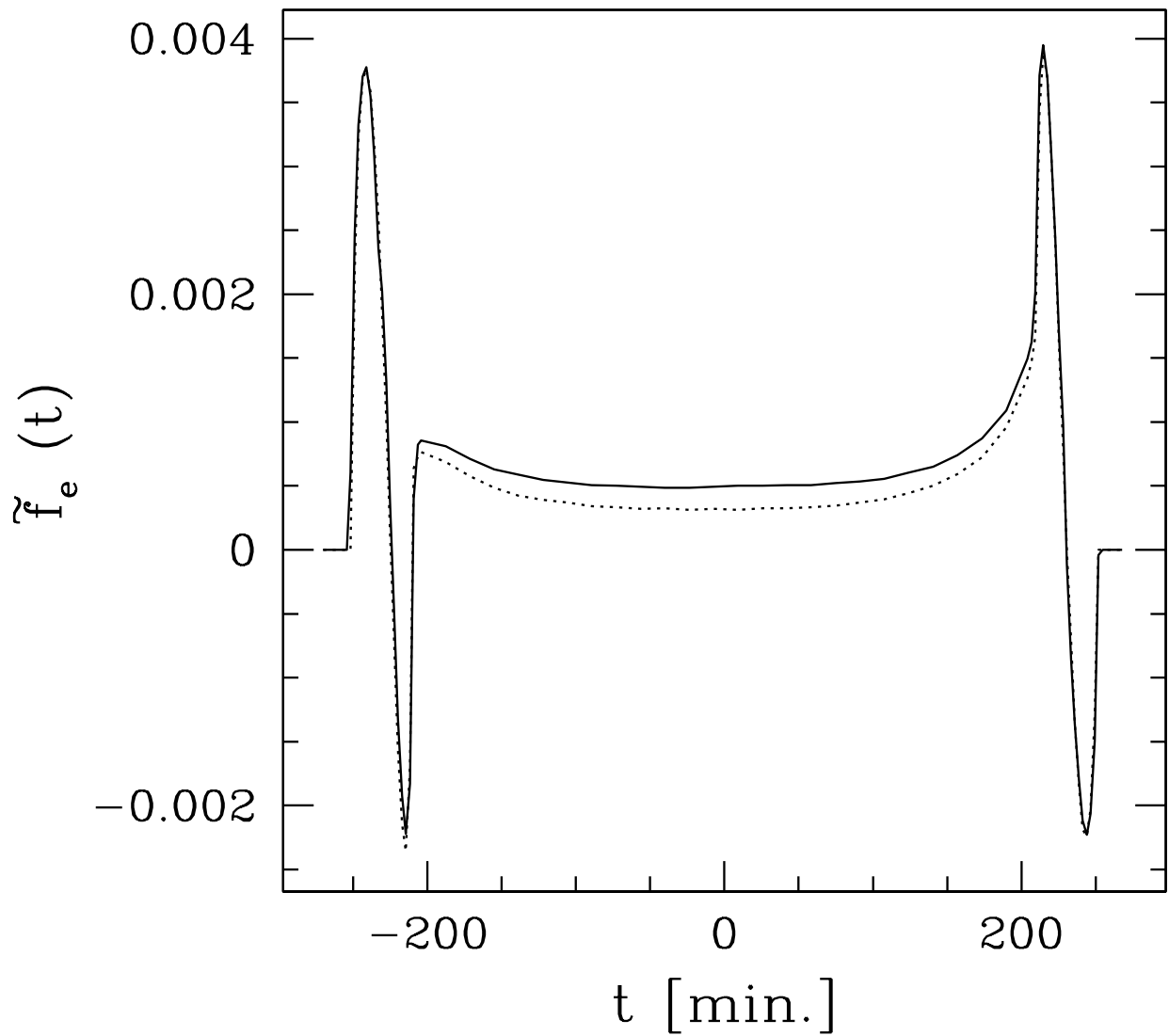,height=7.0in}}
\caption{\label{diff34.Ch} $\tilde f_e$ (equation~(\ref{fetilde})),
the fractional difference between the light deficit from an elliptical
fiducial model and a spherical model with parameters of the spherical
model tuned to minimize this fractional difference. The solid line is
for a model with lensing, and the dotted line is for a model without
lensing. The derived parameters of the spherical planet are $R_0 =
35230$ km, $u_* = 0.8048$, $v_* = -0.227$ for the lensed case (solid
line), and $R_0 = 35270$ km and the same $u_*$ and $v_*$ for the
unlensed case (dotted line). The dotted line is essentially the
fractional difference in the light deficit of an elliptical compared
to a spherical planet of a similar total area. See text for
details.}
\end{figure}

\begin{figure}
\centerline{\psfig{figure=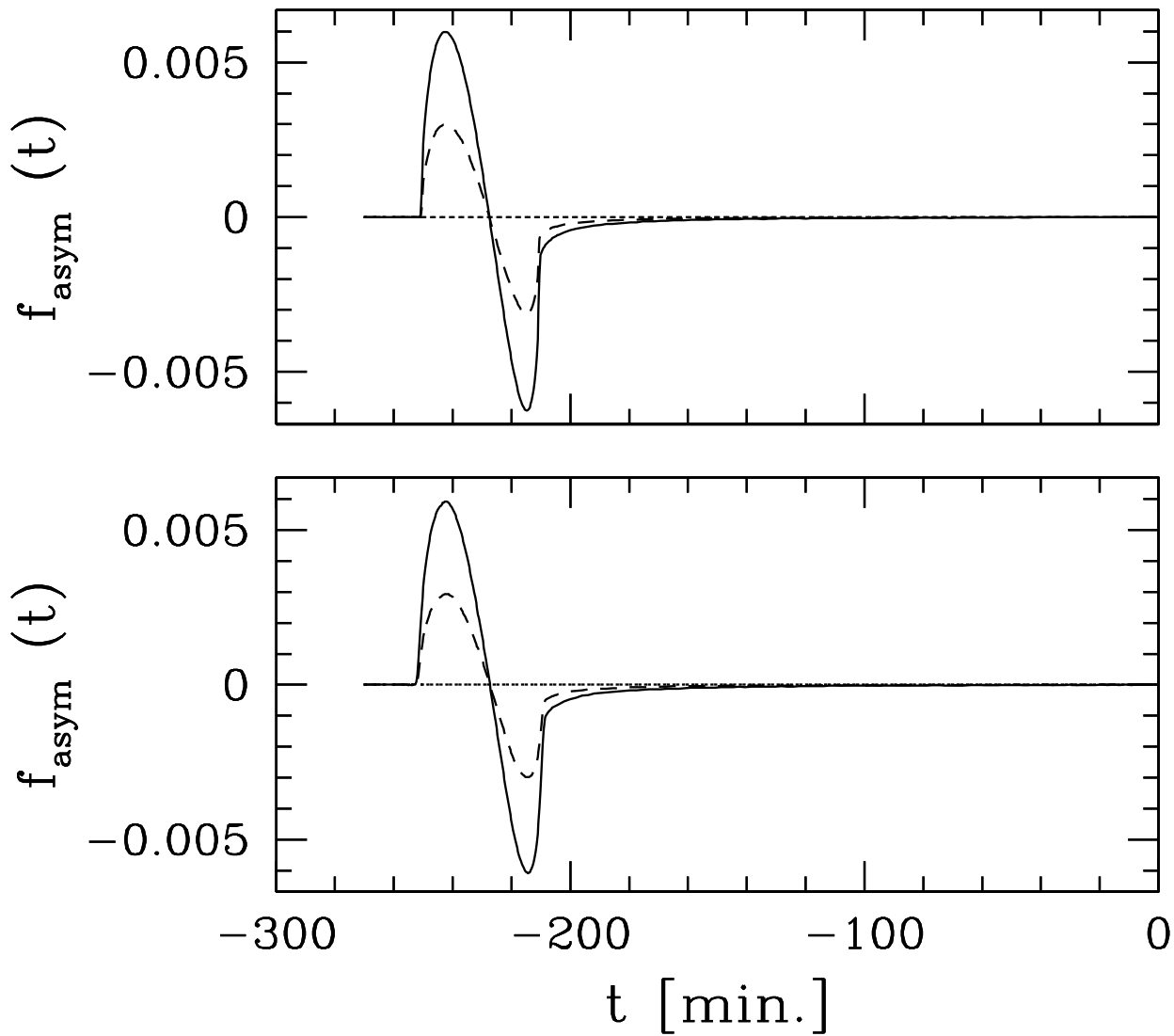,height=7.0in}}
\caption{\label{asymm}
$f_{\rm asym}$ as defined in equation~(\ref{fasym}) with (lower panel) and
without
(upper panel) lensing. The solid line is
for $\epsilon = 0.1$, the dashed line is for $\epsilon = 0.05$, both for
$\gammaimp = 45^0$. The dotted line is for the same $\epsilon$'s
but with
$\gammaimp = 0^0$. In all cases, the rest of the parameters are those
of the fiducial model (solid square in Figure \ref{paramLENS}).
If the curves in the upper and lower panels are plotted together,
they almost completely overlap i.e. light curves with and without
lensing have an almost identical degree of oblateness-induced asymmetry.
}
\end{figure}

\begin{figure}
\centerline{\psfig{figure=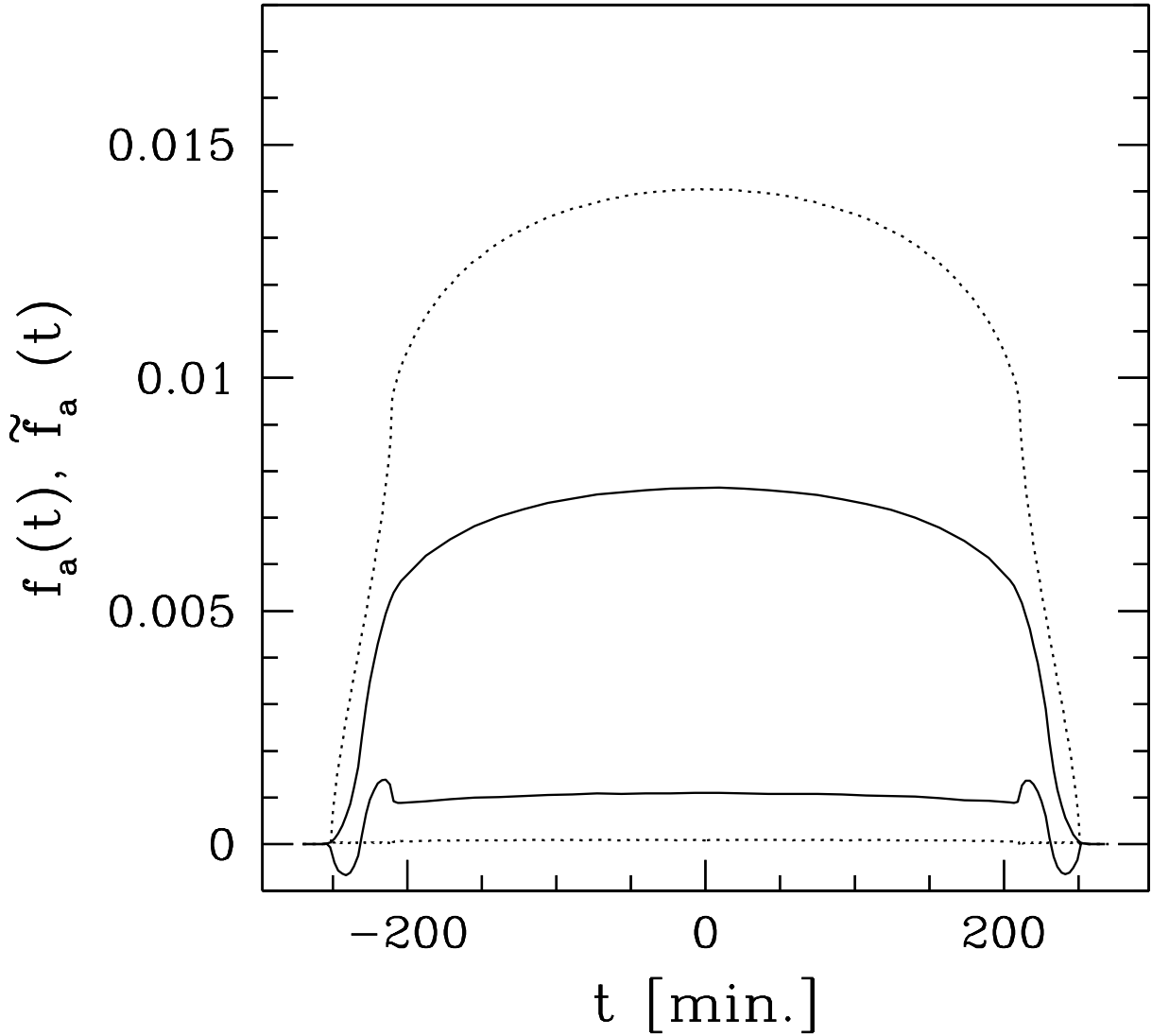,height=7.0in}}
\caption{\label{diffabsorb} The upper solid and dotted lines show $f_a
(t)$ (equation~(\ref{fat})) for cases with and without lensing
respectively.  The lower solid and dotted lines show the corresponding
$\tilde f_a (t)$ (equation~(\ref{fatilde})) which is a minimized
function. The small values of the lower two lines demonstrates that
an appropriate $R_0$ can always be chosen so that the step function
model approximates a more realistic absorption model to high accuracy.
}
\end{figure}

\begin{figure}
\centerline{\psfig{figure=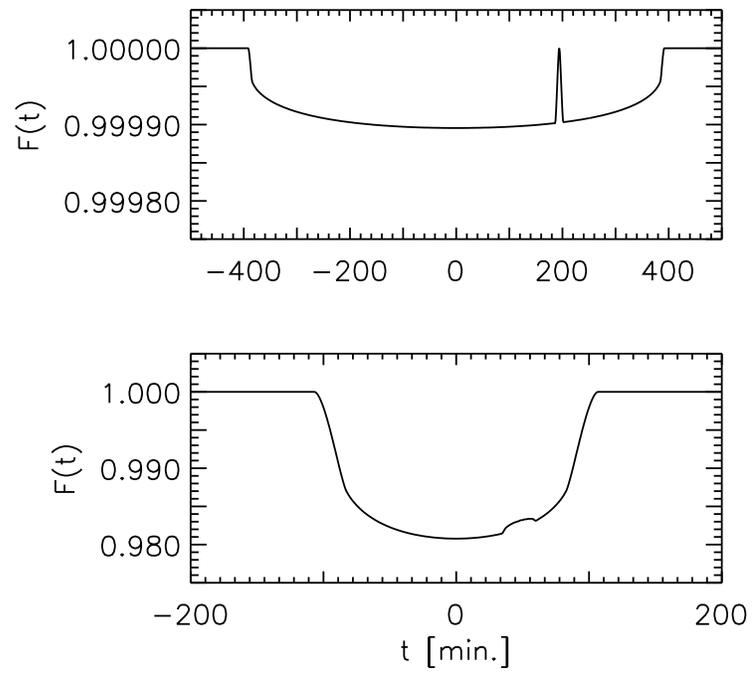,height=4.0in}}
\caption{\label{spot}
The effect of a planet crossing an Earth-sized
star spot on a stellar disk.  Top panel: An Earth-like planet
transit.  Bottom panel: A CEGP-like planet transit.}
\end{figure}

\end{document}